\documentclass[12pt]{article}
\usepackage{amsmath,amssymb,cite,graphicx}

\makeatletter
\@addtoreset{equation}{section}
\makeatother

\def\slashchar#1{\setbox0=\hbox{$#1$}           
   \dimen0=\wd0                                 
   \setbox1=\hbox{/} \dimen1=\wd1               
   \ifdim\dimen0>\dimen1                        
      \rlap{\hbox to \dimen0{\hfil/\hfil}}      
      #1                                        
   \else                                        
      \rlap{\hbox to \dimen1{\hfil$#1$\hfil}}   
      /                                         
   \fi}

\def\tr{\text{tr}}

\def\writecenter#1{
   \rlap{\hbox to 50mm{\hfil#1\hfil}}   
   }

\def\nn{\nonumber}

\def\be{\begin{equation}}
\def\ee{\end{equation}}
\def\ben{\begin{displaymath}}
\def\een{\end{displaymath}}
\def\bea{\begin{eqnarray}}
\def\eea{\end{eqnarray}}
\def\ft#1#2{{\textstyle {\frac{#1}{#2}} }}

\makeatletter \@addtoreset{equation}{section} \makeatother

\def\a{\alpha}

\def\C{\Gamma}

\def\x2{X}

\newcommand{\w}[1]{\\[0.#1cm]}

\def\eq#1{(\ref{#1})}
\def\ft#1#2{{\textstyle{{\scriptstyle #1}\over {\scriptstyle #2}}}}

\def\s2{{\sqrt 2}}

\def\calo{{\cal O}}

\def\tL{{\tilde\Lambda}}
\def\tG{{\tilde G}}

\def\tm{{\tilde\mu}}

\def\bs{\begin{equation}\begin{split}}
\def\es{\end{split}\end{equation}}

\newcommand{\hoch}[1]{$\, ^{#1}$}

\newcommand{\tamphys}{\it George and Cynthia Woods Mitchell  Institute
for Fundamental Physics and Astronomy,\\
Texas A\&M University, College Station, TX 77843, USA}

\newcommand{\auth}{
Roberto Percacci \hoch{\ddagger} and Ergin Sezgin \hoch{\dagger}
}

\begin{document}

\begin{flushright}
\hfill{PI-QG-167}\\
\hfill{MIFP-10-5}
\end{flushright}

\vspace{25pt}

\begin{center}

{\large {\large\bf One Loop Beta Functions in Topologically Massive Gravity }}

\vspace{25pt}
\auth

\vspace{10pt}
\hoch{\dagger}{\tamphys}

\vspace{10pt}

\vspace{10pt}

\hoch{\ddagger}{\it  Perimeter Institute,
31 Caroline St. N, Waterloo, Ontario N2J 2Y5, Canada
\footnote{on leave from SISSA, Trieste, Italy. Supported in part by INFN, Sezione di Trieste.}}

\vspace{40pt}

\underline{ABSTRACT}
\end{center}

We calculate the running of the three coupling constants in cosmological,
topologically massive $3d$ gravity. We find that $\nu$, the dimensionless
coefficient of the Chern-Simons term, has vanishing beta function.
The flow of the cosmological constant and Newton's constant depends on $\nu$,
and for any positive $\nu$ there exist both a trivial and a nontrivial fixed point.

\vspace{15pt}

\pagebreak


%


\tableofcontents

\newpage


\section{Introduction}


Einstein gravity in three dimensions with or without cosmological constant and supplemented with a gravitational Chern-Simons (CS) term has been a subject of many investigations since long \cite{Deser:1981wh}. This theory is of considerable interest, as the presence of the cosmological term makes it possible to have a black hole solution, while the Chern-Simons term is responsible for the presence of a single propagating massive graviton. For the latter reason, this model is referred to as the topologically massive gravity (TMG).

The properties of this theory have been the subject of intense scrutiny.
For a generic value of the CS coupling, either black hole states (if $G<0$) or graviton states
(if $G>0$) will have negative mass.
In \cite{comgrav1} it was observed that in the case $G>0$, if one chooses a critical value of the CS coupling, called the chiral point, the negative mass graviton mode can be confined to propagate only on the boundary, provided that suitable boundary conditions are imposed. The prospects of AdS/CFT duality are enhanced in this framework. There have also been alternative views in which the bulk graviton is maintained but the negative energy black hole solution is viewed as being possibly irrelevant \cite{comgrav2}.

TMG appears to be renormalizable \cite{Deser:1990bj,Keszthelyi:1991ha,Oda:2009im}.
However, not much is known about its renormalization group (RG).
For example, does the chiral point have any special properties?
Another motivation for addressing this issue comes from the asymptotic safety approach to quantum gravity \cite{Weinberg,reviews},
which requires the existence of a fixed point (FP) with finitely many UV attractive directions. Such a nontrivial FP has been found in the Einstein-Hilbert truncation of gravity in any dimension $d>2$ \cite{Lauscher,Fischer, cpr2}, but in $d=3$ this theory does not have propagating degrees of freedom. A nontrivial FP in TMG could provide an example of a three dimensional asymptotically safe theory of gravity with physical degrees of freedom.

Our purpose in this paper is to calculate the one loop beta functions for TMG,
to determine the RG flow of the couplings and to discuss their FPs. We find that many subtleties and technically nontrivial issues arise due to the presence of the Lorentz Chern-Simons. For example, the time honored heat kernel expansion method is not well suited in this case, essentially due to the fact that the eigenvalues of the cubic in derivative wave operator present in the theory is a cubic polynomial in the lowest weight labels, as opposed to being quadratic. There is the delicate matter of how to choose the cut-off scheme to do these computations as well. Our results are best discussed after spelling out all these issues, and are summarized in the Section 7. As we shall see, one of the key results we will obtain is that the dimensionless coefficient, $\nu$,  of the Chern-Simons term has vanishing beta function. Moreover, we find that the flow of the cosmological constant and Newton's constant depends on the values of $\nu$, and for any positive $\nu$ there exist both a trivial and a nontrivial FP. Various aspects of these results are also discussed in Section 7.

This paper is organized as follows. In Section 2, we review briefly the Wilsonian approach to the study of Renormalization Group (RG) equations. In Section 3, we determine the gauge fixed inverse propagator of TMG on a maximally symmetric background. In Section 4, we define the IR cutoff of each spin component separately and we give the harmonic expansion of the fields.
In Section 5, we evaluate the functional traces that arise in the computation of the one-loop beta functions, using the Euler-Maclaurin formula.
The resulting beta functions and their FPs are given
in Section 6. In Section 7, we first summarize our results, and then discuss
various aspects of them a number of subsections.  Four appendices are provided, of which the first two contain useful lemmas and the details of the Euclideanization procedure and harmonic expansions on $S^3$. In Appendix C, we address detail the role of the tadpoles in an off-shell beta function computation, and in Appendix D we give a heat kernel evaluation of the beta functions of pure gravity with cosmological constant, which provide a useful check of the techniques used in Section 5.


\section{Wilsonian Method For Computing the RGE}


There is no single approach to the study of Renormalization Group (RG) equations. Here, we shall adopt a Wilsonian method. The central lesson of Wilson's analysis of QFT is that the ``effective'' (as in ``effective field theory'') action describing physical phenomena at a momentum scale $k$ can be thought of as the result of having integrated out all fluctuations of the field with momenta larger than $k$. Since $k$ can be regarded as the lower limit of some functional integration, we will usually refer to it as the infrared cutoff. The dependence of the ``effective'' action on $k$ is the Wilsonian RG flow
\footnote{At this general level of discussion, it is not necessary to specify the physical meaning of $k$: for each application of the theory one will have to identify the physically relevant variable acting as $k$. In scattering experiments $k$ is usually identified with some external momentum. See \cite{Hewett:2007st} for a discussion of this choice in concrete applications to gravity.}.

There are several ways of implementing this idea in practice,
resulting in several forms of the RG equation  \cite{wegner}. In the specific
implementation that we shall use \cite{Wetterich:1992yh}, instead of introducing a sharp
cutoff in the functional integral, we suppress the contribution of the field
modes with momenta lower than $k$. This is obtained by modifying the low momentum end of the propagator, and leaving all the interactions unaffected.
We describe first the general idea, and comment on its application to gravity later.
We start from a bare action $S[\phi]$ for some fields $\phi$,
and we add to it a suppression term $\Delta S_{k}[\phi]$ that is quadratic in the field. In flat space this term can be written simply in momentum space.
In order to have a procedure that works in an arbitrary curved spacetime we choose a suitable differential operator $\calo$ whose eigenfunctions $\varphi_n$, defined by $\calo\varphi_n=\lambda_n\varphi_n$, can be taken as a basis in the functional space we integrate over:
\begin{equation}
\phi(x)=\sum_n \tilde\phi_n\varphi_n(x)\ ,
\end{equation}
where $\tilde\phi_n$ are generalized ``Fourier components'' of the
field. (We will use a notation that is suitable for an operator with
a discrete spectrum.) Then, the additional term can be written in
either of the following forms:
\begin{equation}
\label{cutoffterm}
\Delta S_{k}[\phi]=
\frac{1}{2}\int dx\,\phi(x)\, R_{k}(\calo)\,\phi(x)
=\frac{1}{2}\sum_n \tilde\phi_n^2 R_k(\lambda_n)\,.
\end{equation}
The kernel $R_{k}(\calo)$ will also be called ``the cutoff''. It is
arbitrary, except for the general requirements that $R_k(z)$ should be
a monotonically decreasing function both in $z$ and $k$,
that $R_k(z)\to 0$ for $z \gg k$ and $R_k(z)\not= 0$ for $z \ll k$.
These conditions are enough to guarantee that the contribution to
the functional integral of field modes $\tilde\phi_n$
corresponding to eigenvalues $\lambda_n\ll k^2$ are
suppressed, while the contribution of field modes corresponding to
eigenvalues $\lambda_n\gg k^2$ are unaffected.
We will further require that  $R_k(z)\to k^2$ for $k\to 0$.
We define a $k$-dependent generating functional of connected Green functions by
\begin{equation}
\label{ge}
e^{-W_{k}\left[J\right]}= \int D\phi\exp\left\{ -S[\phi]-\Delta
S_{k}[\phi]-\int dx\, J\phi\right\}
\end{equation}
and a modified $k$-dependent Legendre transform
\begin{equation}
\label{legendre}
\Gamma_{k}[\phi]=W_{k}\left[J\right]-\int dx\, J\phi-\Delta
S_{k}[\phi]\,,
\end{equation}
where $\Delta S_{k}[\phi]$ has been subtracted. The functional
$\Gamma_k$ is sometimes called the ``effective  average action'',
because it is closely related to the effective action for fields
that have been averaged over volumes of order $k^{-d}$ ($d$ being
the dimension of spacetime). The ``classical fields'' $\delta
W_{k}/\delta J$ are denoted again $\phi$ for notational simplicity.
In the limit $k\to 0$ this functional tends to
the usual effective action $\Gamma[\phi]$, the generating functional
of one-particle irreducible Green functions.
It is similar in spirit to the Wilsonian effective action,
but differs from it in the details of the implementation.

The average effective action $\Gamma_{k}[\phi]$, used at tree level,
should give an accurate description of processes occurring at momentum
scales of order $k$. In the spirit of effective field theories,
one assumes that $\Gamma_k$ exists and admits a derivative expansion of the form
\begin{equation}
\label{expansion}
\Gamma_{k}(\phi,g_{i})=
\sum_{n=0}^\infty\sum_{i}g_{i}^{(n)}(k)\calo_{i}^{(n)}\left(\phi\right)\ ,
\end{equation}
where $g_{i}^{(n)}(k)$ are coupling constants and $\calo_{i}^{(n)}$
are all possible operators constructed with the field $\phi$ and
$n$ derivatives, which are compatible with the
symmetries of the theory. The index $i$ is used here to label
different operators with the same number of derivatives.
At one loop, $\Gamma_k$ is given by
\be
\label{ansatz}
\Gamma^{(1)}_k=S+\frac{1}{2}{\rm Tr}\,{\rm log}\left(S^{(2)}+R_k\right)
\ee
where $S^{(2)}$ denotes the second variation of the bare action.
The only dependence on $k$ is contained in the cutoff term $R_k$
and therefore
\begin{equation}
\label{oneloop}
k\frac{d\Gamma^{(1)}_k}{dk} = \frac{1}{2}\mathrm{Tr}
\left(S^{(2)}+R_k\right)^{-1}k\frac{dR_k}{dk}\ .
\end{equation}
Note that due to the properties of $R_k(z)$, the factor $k\frac{dR_k}{dk}$
goes to zero for $z>k^2$. Thus, even though $R_k(z)$ is introduced in the functional integral as an infrared cutoff, in the evaluation of the beta functions it acts effectively as UV cutoff.
One can extract from this functional equation the one loop beta functions,
by making an ansatz for $\Gamma^{(1)}_k$ of the form \eq{expansion},
possibly containing a finite number of terms,
inserting it in the left hand side of \eq{oneloop},
expanding the right hand side and comparing term by term.

When applied to scalar, fermion and gauge field theories this method reproduces the well known one loop beta functions. In this paper we shall apply this technique to TMG. In general, in applying Wilsonian ideas to gravity one has to confront the fact that the definition of a cutoff generally makes use of a metric and since the metric is now to be treated as a dynamical field it is less clear what one means by cutoff. We will follow \cite{reuter1} and use the background field method, thereby effectively replacing the dynamical metric $g_{\mu\nu}$ by a spin two field $h_{\mu\nu}$ propagating in a fixed but unspecified background $\bar g_{\mu\nu}$. In addition to appearing in the gauge fixing, the background metric can be used to unambiguously distinguish what is meant by long and short distances, and hence low and high energies. In practice, it allows us to write a cutoff term that is purely quadratic in the quantum field, as required in the preceding discussion. The effective average action is then a functional of two fields $\Gamma_k(\bar g_{\mu\nu},h_{\mu\nu})$; in practice, aside from the gauge fixing and cutoff terms, which are quadratic in $h_{\mu\nu}$, we will only need the restriction of the functional $\Gamma_k$ to the space where $\langle h_{\mu\nu}\rangle=0$. This is sufficient to extract beta functions for the couplings we are interested in.


\section{The Quadratic Gauge Fixed Action for TMG}


Topological massive gravity is described the action
\be
S =  Z \int d^3 x \sqrt{-g} \left( R  -2\Lambda
+\ft1{2 \mu}\,\varepsilon^{\lambda\mu\nu} \C_{\lambda\sigma}^\rho \left( \partial_\mu \C_{\nu\rho}^\sigma +\ft23 \C_{\mu\tau}^\sigma \C_{\nu\rho}^\tau \right)\right)
   \ , \label{1}
\ee
where $Z={1\over 16 \pi G}$
\footnote{We use the conventions
$R_{\mu\nu}{}^\rho{}_\sigma=\partial_\mu\Gamma_\nu{}^\rho{}_\sigma
-\partial_\nu\Gamma_\mu{}^\rho{}_\sigma
+\Gamma_\mu{}^\rho{}_\tau\Gamma_\nu{}^\tau{}_\sigma
-\Gamma_\nu{}^\rho{}_\tau\Gamma_\mu{}^\tau{}_\sigma$
and $R_{\mu\nu}=R_{\rho\mu}{}^\rho{}_\nu$.}.
At this point we need not specify the signs of the couplings $G,\Lambda$ and $\mu$.
It will be useful to define the dimensionless combinations
\be
\label{dimless}
\nu=\mu G\ ;\qquad
\tau=\Lambda G^2\ ;\qquad
\phi=\mu/\sqrt{|\Lambda|}\ .
\ee
Note in particular that $\frac{1}{32\pi\nu}$ is the coefficient of the CS term.
We will expand around background metric ${\bar g}_{\mu\nu}$ as
$g_{\mu\nu} = {\bar g}_{\mu\nu} +h_{\mu\nu}$.
Next, we add to the action the gauge fixing term
\be
S_{GF} =  -\frac{Z}{2\a}\int d^3 x \sqrt{-\bar g} \chi_\mu {\bar g}^{\mu\nu} \chi_\nu\ ,
\ee
where
\be
\label{gf}
\chi_\nu =\partial_\mu h^{\mu\nu} -\frac{\beta + 1}{4} \partial_\nu h\ .
\ee
Note that $\beta=1$ corresponds to the familiar de Donder gauge.
The ghost action corresponding to the gauge \eq{gf} is given by
\be
S_{gh} = -\int d^3 x \sqrt {-g} {\bar C}^\mu \left( \delta_\mu^\nu \Box +\frac{1-\beta}{2} \nabla_\mu\nabla^\nu +R_\mu{}^\nu  \right) C_\nu\ .
\label{gh}
\ee
The one loop average effective action $\Gamma_k^{(1)}$ contains terms of the
same form as $S+S_{GF}+S_{gh}$, except that all the couplings in $S$ are now to be interpreted as renormalized couplings
\footnote{We allow only the couplings in $S$ to run, whereas the gauge fixing parameters will be kept fixed.}.
Their beta functions will be obtained from equation \eq{oneloop}, which for this theory has the form
\be
\label{gravoneloop}
k\frac{d\Gamma^{(1)}_k}{dk} =
\frac{1}{2}\mathrm{Tr}\left( \frac{\delta^2\left(S+S_{GF}\right)}{\delta h_{\mu\nu} \delta h_{\rho\sigma}} +{\cal R}^{\mu\nu\rho\sigma}\right)^{-1}k \frac{d {\cal R}_{\rho\sigma\mu\nu}}{dk}
- \mathrm{Tr}\left( \frac{\delta^2 S_{gh}}{\delta {\bar C}^\mu \delta C_\nu}
+{\cal R}_\mu{}^\nu  \right)^{-1}  k \frac{d {\cal R}^\mu{}_\nu}{dk}\ .
\ee

Expanding in powers of $h_{\mu\nu}$, and discarding total derivative terms,
the quadratic part of the action is given by
%
%
\bea
S^{(2)} + S_{GF} &=&  \frac{1}{4}Z \int d^3 x \sqrt {-g}
\Bigg[ h_{\mu\nu} \left(\Box -\frac{2R}{3} +2\Lambda \right) h^{\mu\nu}
+\frac{2(1-\alpha)}{\alpha} h_{\mu\nu} \nabla^\mu\nabla_\rho h^{\rho\nu}
\nn\w2
&& +\left(2-\frac{\beta+1}{\alpha} \right) h \nabla^\mu\nabla^\nu h_{\mu\nu}
-\left(1-\frac{(\beta+1)^2}{8\alpha}\right) h\Box h
+ \frac16 h (R-6\Lambda) h
\nn\w2
&& +\frac1{\mu} \varepsilon^{\lambda\mu\nu} h_{\lambda\sigma} \left( \nabla_\mu \left(\Box-\frac{R}{3}\right) h^\sigma{}_\nu -\nabla_\mu\nabla^\sigma\nabla^\rho h_{\rho\nu}\right)
\Bigg]\ .
\label{a2}
\eea
The raising and lowering of indices is understood to be with the background metric, on which we drop the bar for notational simplicity. For our purposes it will be sufficient to consider maximally symmetric backgrounds,
for which
\be
R_{\mu\nu\rho\sigma}= \frac{R}{6} (g_{\mu\rho}g_{\nu\sigma} -g_{\mu\sigma}g_{\nu\rho})\ , \qquad R_{\mu\nu}= \frac{R}{3} g_{\mu\nu}\ .
\label{sos}
\ee
Here $R=\pm 6/\ell^2$, with the $+$ sign for de Sitter, and the $-$ sign for anti-de Sitter space, and $\ell$ is the ``radius''. It is important to note that, in computing the gauge fixed quadratic action and the beta functions, we do not use the equation of motion which for this class of metrics would imply the relation $\ell= 1/\sqrt{|\Lambda|}$. As we shall see later, this is necessary to compute the beta functions for $\Lambda$ and $G$ separately.

In order to achieve partial diagonalization of the inverse propagator we decompose the quantum fluctuation $h_{\mu\nu}$ into irreducible parts:
\be
h_{\mu\nu}=h^T_{\mu\nu}+\nabla_\mu\xi_\nu+\nabla_\nu\xi_\mu
+\nabla_\nu\nabla_\nu\sigma-{1\over3}g_{\mu\nu}\Box\sigma+{1\over3} g_{\mu\nu}h\ ,
\label{decomp}
\ee
where $h=g^{\mu\nu}h_{\mu\nu}$, $\xi_\mu$ satisfies
$\nabla^\lambda\xi_\lambda=0$ and
$h^T_{\mu\nu}$ satisfies $g^{\mu\nu}h^T_{\mu\nu}=0$
and $\nabla^\lambda h^T_{\lambda\nu}=0$. It is important to note that in passing from a path integral over the field $h_{\mu\nu}$ to one over the fields $(h_{\mu\nu}^T, \xi, \sigma,h)$, one has to take into account the appropriate Jacobian factors.
These Jacobians will be exactly canceled by other Jacobians arising from suitable redefinitions of $\xi$ and $\sigma$, as we will see later (see \eq{redefs}).

Inserting \eq{decomp} in \eq{a2}, and using the lemmas in Appendix A, we find
%
%
\bea
\label{master2}
S^{(2)}+S_{GF}\!\!\! &=&\!\!\! {1\over 4}Z_k \int d^3 x \sqrt{-g} \Biggl\{
h^T_{\mu\nu}\left(\Box-{2\over3}R+2\Lambda_k\right)h^{T\mu\nu}
+\frac1{\mu} h_{\lambda\sigma}^{T} \left(\Box-\frac{R}3\right) \varepsilon^{\lambda\mu\nu} \nabla_\mu h^T_\nu{}^\sigma
\nn\w2
&&\!\!\!\!\!\!\!\!-{2\over \alpha} \xi_\mu \left(\Box +\frac{R}3\right)
\left(\Box+{1-\alpha\over 3}R+2\alpha\Lambda_k\right)\xi^\mu
\nn\w2
&&\!\!\!\!\!\!\!\! +\frac{2(4-\alpha)}{9\alpha} \sigma \Box \left(\Box +\frac{R}2\right)
\left(\Box +{2-\alpha\over 4-\alpha}R
+{6\alpha\over4-\alpha}\Lambda_k\right)\sigma
\w2
&&\!\!\!\!\!\!\!\!+ \frac{2(2\alpha-3\beta+1)}{9\alpha}\sigma\Box\left({\Box+{R\over2}}\right)\,h
-h\left(\frac{6\beta-9\beta^2+16\alpha-1}{72\alpha}\Box
+\frac{R}{18}+\frac{\Lambda_k}3 \right)h\Biggr]\Biggr\}\ .
\nonumber
\eea
Note the factor $1/4$ instead of the traditional $1/2$.
The additional, irrelevant factor of $1/2$ has been extracted in order to have an operator where the coefficient of $\Box$ acting on $h^T$ is one.
Note that the CS term affects only the propagation of $h_{\mu\nu}^{T}$.
Similarly we decompose the ghost field as
\be
C_\mu = V_\mu + \partial_\mu S\ ,
\label{ghostdecomp}
\ee
where $\nabla_\mu V^\mu=0$, and similarly for $\bar C^\mu$.
This leads to
\be
S^{(2)}_{\rm ghost} = -\int d^3 x \sqrt {-g} \left[ {\bar V}^{\mu} \left( \Box +\frac{R}3 \right) V_\mu
-\frac{3-\beta}{2}  {\bar S}\, \Box \left(\Box +\frac{4}{3(3-\beta)} R\right) S\right]\ .
\label{ghost}
\ee
Changing ghost variables to $(V_\mu,S)$ again produces a Jacobian
which will be canceled by another Jacobians coming from a redefinition of $S$ (see \eq{redefs}).
%
%


\section{The Cutoff}


In this section we define the cutoff and we set up the calculation
of the traces in \eq{gravoneloop} by means of harmonic expansions
on $S^3$, viewed as Euclideanized de Sitter space.
From here on we will work in the ``diagonal'' gauge in which $\sigma-h$ couplings vanish. This fixes
\be
\beta=\frac{2\alpha+1}{3}\ .
\label{ab}
\ee
Furthermore we make the field redefinitions
\be
\sqrt{\Box +\frac{R}3 }\ \xi_\mu  = {\hat \xi}_\mu\ ;\qquad
\sqrt{\Box\left(\Box +\frac{R}2\right)}\ \sigma = {\hat\sigma}\ ;\qquad
{\sqrt \Box} S = {\hat S}\ ,
\label{redefs}
\ee
whose Jacobian determinants cancel those coming from \eq{decomp}
and \eq{ghostdecomp} \cite{Fradkin:1982bd,Dou}. Then the action \eq{master2} becomes
\be
S^{(2)}+S_{GH} ={1\over 4}Z_k \int d^3 x \sqrt{-g}\left[ h^{TT\mu\nu } \Delta_{2\mu\nu}{}^{\rho\sigma} h_{\rho\sigma}^{TT} + c_1 {\hat\xi}^\mu \Delta_{1\mu}{}^\nu {\hat\xi}_\nu + c_\sigma {\hat\sigma}\Delta_\sigma {\hat\sigma} + c_h h \Delta_h h \right]\ ,
\ee
where we have defined the operators
\bea
\label{ops}
\Delta_{2\mu\nu}{}^{\rho\sigma} &=& \left(\Box -\frac{2R}{3} + 2\Lambda\right) \delta_{(\mu}^{(\rho} \delta_{\nu)}{}^{\sigma)} +\frac{1}{\mu} \varepsilon_{(\mu}{}^{\lambda(\rho}\delta_{\nu)}^{\sigma)}  \nabla_\lambda\sqrt{\Box(\Box -\frac{R}{3})} \ ,
\nn\w2
\Delta_{1\mu}{}^\nu &=& \Box +\frac{1-\alpha}{3} R +2\alpha\Lambda\ ,
\nn\w2
\Delta_\sigma &=& \Box +\frac{2-\alpha}{4-\alpha} R +\frac{6\alpha\Lambda}{4-\alpha} \ ,
\nn\w2
\Delta_h &=& \Box +\frac{R}{4-\alpha} +\frac{6\Lambda}{4-\alpha}\ ,
\eea
and coefficients
\be
c_1 = -\frac{2}{\alpha}\ ,\qquad
c_\sigma = \frac{2(4-\alpha)}{9\alpha}\ ,\qquad
c_h = -\frac{4-\alpha}{18}\ .
\ee
Similarly, the ghost action \eq{ghost} in the diagonal gauge becomes
\be
\label{gh2}
 S^{(2)}_{\rm ghost} = -\int d^3 x \sqrt {-g} \left[ {\bar V}^{\mu} \Delta_{V\mu}{}^\nu V_\nu
+c_S\bar{\hat S} \left(\Box +\frac2{4-\alpha} R\right) {\hat S} \right]\ ,
\ee
where $c_S=(\alpha-4)/3$.
It would be tempting to choose $\alpha=1$, which corresponds to the de Donder gauge and simplifies the inverse propagator, but we shall not do so in order to probe the $\alpha$-dependence of our  results. We shall come back to this issue later.
We note that the gauge $\alpha=4$ which has been used in \cite{comgrav1}
is not singular, as can be seen from \eq{master2}.
However, in this gauge the $\sigma^2,h^2,{\hat S}^2$ terms do not contain $\Box$
and the method we shall use below to compute the beta functions will turn out to be not suitable to deal with this case. Therefore, in the rest of this paper we will assume that $\alpha\ne 4$.

Now we are ready to discuss the choice of the cutoff.
For each spin component we choose the cutoff to be a function of the
corresponding operator given in \eq{ops}.
Then, defining the gauge fixed inverse propagator
\be
{\cal O}=
Z
\left(
  \begin{array}{cccc}
    \Delta_2 &  &  &  \\
     & c_1\Delta_1 &  &  \\
     &  & c_\sigma\Delta_\sigma &  \\
     &  &  & c_h\Delta_h \\
  \end{array}
\right)
\ee
we choose the cutoffs to have the following forms
\be
{\cal R}_k=
Z
\left(
  \begin{array}{cccc}
    R_k(\Delta_2) &  &  &  \\
     & c_1R_k(\Delta_1) &  &  \\
     &  & c_\sigma R_k(\Delta_\sigma) &  \\
     &  &  & c_h R_k(\Delta_h) \\
  \end{array}
\right)
\ee
where $R_k(z)$ is a real profile function with the properties described in Section 1.
Then the denominator of the first term in \eq{gravoneloop} is
\be
{\cal O}+{\cal R}_k=
Z
\left(
  \begin{array}{cccc}
    P_k(\Delta_2) &  &  &  \\
     & c_1 P_k(\Delta_1) &  &  \\
     &  & c_\sigma P_k(\Delta_\sigma) &  \\
     &  &  & c_h P_k(\Delta_h) \\
  \end{array}
\right)
\ee
where we have defined the function $P_k(z)=z+R_k(z)$.
The bare couplings appearing in the second variation
of the action do not depend on $k$, so in the numerator of \eq{gravoneloop} we use the following formula
\be
\label{rdot}
\frac{d}{dt}\left(Zc_i R_k(\Delta_i)\right)=
Zc_i\partial_t R_k(\Delta_i)\ .
\ee
Then the overall factors $Z$ and $c_i$ cancel between numerator and denominator and \eq{gravoneloop} reduces to
\bea
\label{erge3}
k\frac{d\Gamma_k}{dk} &=&
\frac{1}{2}\left[\mathrm{Tr}_2 W(\Delta_2)
+\mathrm{Tr}_1 W(\Delta_1)
+\mathrm{Tr}_0 W(\Delta_\sigma)
+\mathrm{Tr}_0 W(\Delta_h)\right]
\nonumber\\&&
- \left[\mathrm{Tr}_1W(\Delta_V)+\mathrm{Tr}_0W(\Delta_S)\right]
\ ,
\eea
where we have defined the function $W(z)=\frac{\partial_t R_k}{P_k}$.

The traces are given by the sum of the functions over all the eigenvalues.
Such sums can be performed after Euclidean continuation to a three sphere.
This is described in Appendix B. See Section 5 for further comments.

We must now specify the cutoff profile $R_k$.
In order to be able to perform the summations in closed form we will use
the ``optimized'' cutoff $R_k(z)=(k^2-|z|)\theta(k^2-|z|)$ \cite{Litim:2001up}.
Note the absolute value, which is normally not used because the operators
one normally considers are positive.
Insofar as the role of the cutoff is to limit the sum to a finite number of modes,
this definition will work also if the operator is negative.
We will discuss this point at some length in the next section.
We have
$\partial_t R_k(z)=2k^2\theta(k^2-|z|)$ and for $z<k^2$, $P_k(z)=k^2$,
so that $W(z)=2\theta(k^2-|z|)$.
Dividing numerator and denominator by $k^2$, they are given by
\bea
\label{sums}
k\frac{d\Gamma_k}{dk} &=&
\sum_{\pm}\sum_{n} m_n^{T\pm}\theta(1-|\tilde\lambda_n^{T\pm}|)
+\sum_{n}m_n^\xi\theta(1-\tilde\lambda_n^\xi)
\notag\\
&&+\sum_{n}m_n^\sigma\theta(1-\tilde\lambda_n^\sigma)
+\sum_{n}m_n^h\theta(1-\tilde\lambda_n^h)
\notag\\
&&- 2\sum_n  m_n^V\, \theta(1-\tilde\lambda_n^V)
- 2\sum_n m_n^S\, \theta(1-\tilde\lambda_n^S)
\ .
\eea
Here $\tilde\lambda_n^{(i)}=\lambda_n^{(i)}/k^2$ are the distinct dimensionless
eigenvalues of the Euclideanized operator $\Delta_i$, and $m_n^{(i)}$ their multiplicities.
Using the results of Appendix B, the eigenvalues are given by
\bea
\lambda_n^{T\pm} &=& \frac{R}{6}(n^2+2n+2)  -2\Lambda \pm \frac{1}{\mu}\left(\frac{R}{6}\right)^{3/2} n(n+1)(n+2)\ ,\quad n\ge 2\ ,
\nn\w2
\lambda^\xi_n &=& \frac{R}{6}\left(n^2+2n-3+2\alpha \right) -2\alpha\Lambda\ , \quad n\ge 2\ ,
\nn\w2
\lambda^\sigma_n  &=& \frac{R}{6}
\left(n^2+2n-\frac{6(2-\alpha)}{4-\alpha}\right) -\frac{6\alpha\Lambda}{4-\alpha}  \ , \quad n\ge 2\ ,
\nn\w2
\lambda^h_n &=& \frac{R}{6}\left( n^2 + 2n - \frac{6}{4-\alpha}\right) -\frac{6\Lambda}{4-\alpha}\ ,\quad n\ge 0\ ,
\eea
and the respective multiplicities are
\bea
\label{multiplicities}
m^{T+}_n= m^{T-}_n &=& n^2+2n-3\ ,\nn\w2
m^\xi_n=m_n^V&=& 2(n^2+2n)\ ,\nn\w2
m^\sigma_n=m_n^h=m_n^S&=& n^2+2n+1\ .
\eea
We see that the effect of the cutoff function is simply to terminate each
sum over eigenvalues at some maximal value $n_{\rm max}$.
In the next section we discuss the way in which one can
evaluate these finite sums, emphasizing the subtleties that arise
in the spin two sector.


\section{Evaluation of the functional Traces}


The most familiar way of evaluating the asymptotic expansion of functional traces
is the method of the heat kernel.
However, in the presence of the $\mu$--dependent terms in the trace over the spin 2 fields,
the heat kernel coefficients are not known.
Since on the other hand for a maximally symmetric background the eigenvalues are known explicitly, we will directly evaluate the expansion of the traces using the Euler-Maclaurin formula. In the present context, where the function to be traced and all its derivatives vanish at $+\infty$, it says that
\be
\sum_{n=n_0}^\infty F(n)=\int_{n_0}^\infty dx\,F(x)+\frac{1}{2}F(n_0)-\frac{B_2}{2!}F'(n_0)-\frac{B_4}{4!}F'''(n_0)+R
\label{ml}
\ee
where $B_n$ are the Bernoulli numbers and $R$ is a remainder.
For the evaluation of the beta functions of interest, the terms with three or more derivatives of $F$ give no contribution, so it is enough to include
the term containing $B_2=1/6$ and the remainder can be neglected.

The evaluation of the sums can be done using algebraic manipulation software.
There are some subtleties that need to be emphasized.
For each type of field, the effect of the step function is that
the integral extends up to a finite value of $n$.
Thus, in any spin sector in \eq{sums} we have to evaluate, schematically
\be
\int_{n_0}^\infty dx\, m(x) \theta (1-\tilde\lambda(x)) =\int_{n_0}^{n_{\rm max}} dx\, m(x)\ ,
\label{nmax}
\ee
where $n_{\rm max}$ is a (real and positive) root of the equation
\footnote{In general $n_{\rm max}$ is not an integer. One can think that the sums can be evaluated by terminating them at the largest integer $n\le n_{\rm max}$. However, it is easy to check that his procedure applied, for example, to the evaluation of the known heat kernel coefficients will not give the right answer.}
\be
\label{rootz}
\tilde\lambda(x)-1=0\ .
\ee
When the eigenvalues are quadratic in $n$, the equation \eq{rootz} has two roots.
Since the cutoff must be positive, it is always clear which one to choose.
For $\alpha<4$, which is the case for the gauges we use,
the trace $h$ has a negative definite kinetic term.
This is the well known problem of the indefiniteness of the
gravitational action.
In the evaluation of the sums \eq{sums} this sign has no effect.
One can choose $n_{\rm max}$ as for the other components and
this does not lead to any pathology in the flow equation
(see \cite{reuter1} for a discussion).

Things are more complicated in the presence of the CS term.
The eigenvalues of the TT field contain a term which is cubic in $n$,
and this term occurs with opposite signs in the two chirality sectors.
If we tried to define directly the one loop effective action,
this would lead to convergence problems similar to those discussed
in \cite{Witten:2010cx}. For us, convergence should never be a problem because we are only interested in the $k$-derivative of the one loop effective action, which is finite due to the cutoff term. Still, one has to deal with the fact that the $+$ and $-$ eigenvalues have different signs (at least for sufficiently large $n$) and grow at different rates.

Since for $\tilde\lambda^{T\pm}$ equation \eq{rootz} is cubic, one has to choose one among its three roots
\be
r-s+t\ ;\qquad r+e^{i\pi/3}s-e^{-i\pi/3}t\ ;\qquad r+e^{-i\pi/3}s-e^{i\pi/3}t\ ,
\ee
where $r$, $s$, $t$ depend on $\tilde R=R/k^2$, $\tilde\Lambda=\Lambda/k^2$ and $\tilde\mu=\mu/k$.
Superficially the first root seems to be real and the others complex, so
one could try to define the cutoff using the first root.
However, the expressions $s$ and $t$ actually involve cubic roots
and there are regions in parameter space where these roots develop
imaginary parts. In fact, there is no root which is real for all parameter values.

%
\begin{figure}
\begin{center}
{\resizebox{1\columnwidth}{!}
{\includegraphics{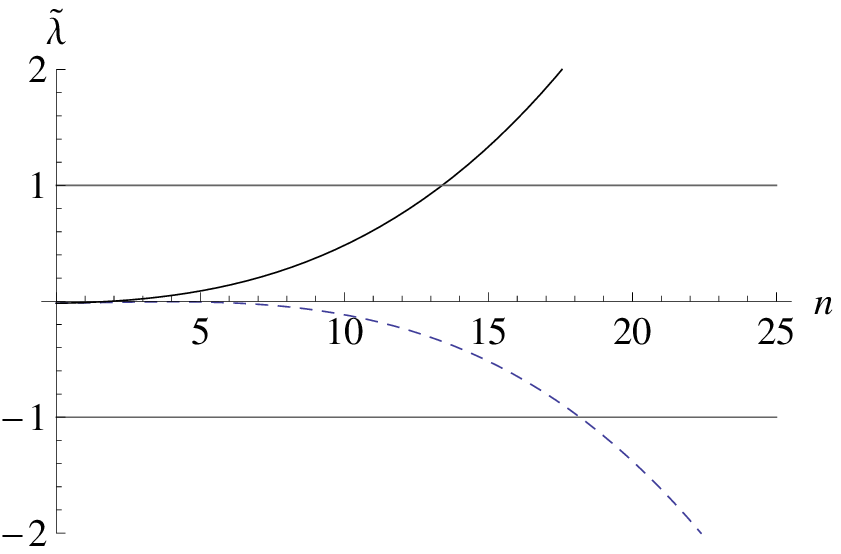}
\includegraphics{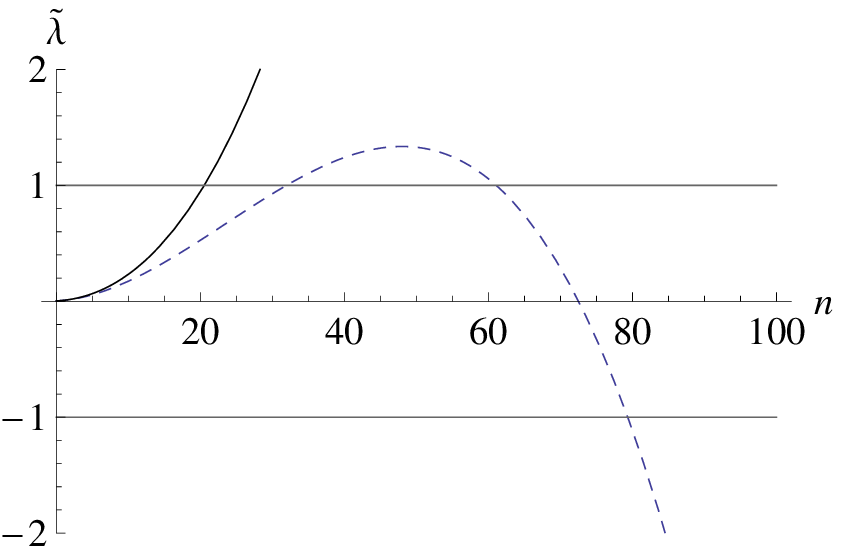}}
\caption{\small The eigenvalues $\tilde\lambda_n^{T+}$ (solid curve)
and $\tilde\lambda_n^{T-}$ (dashed curve) as functions of $n$,
for $\tilde R=\tilde\Lambda=0.01$.
Left panel: large $\tilde\mu$ regime (here $\tilde\mu=3$).
Right panel: small $\tilde\mu$ regime (here $\tilde\mu=0.3$).
}}
\end{center}
\end{figure}
One can understand this better by considering the plot
of the functions $\tilde\lambda^{T\pm}(x)$ for $x>0$, see Figures 1,2 and 3.
Consider first the case when $\tilde\mu>0$.
The function $\tilde\lambda^{T+}(x)$ is monotonically increasing,
and for this function equation \eq{rootz} has a single real root.
Thus there is no ambiguity for the positive chirality modes.
In the case of the negative chirality modes, however,
for small $x$ the function $\tilde\lambda^{T-}(x)$ initially grows with $x$,
until the cubic term prevails; hereafter it decreases.
When $\tilde R$ and $\tilde\Lambda$ can be neglected,
which is the situation we are interested in,
the maximum is equal to $4\tilde\mu^2/27$.
Thus if $\tilde\mu>\sqrt{27/4}$ equation \eq{rootz} has two positive roots.
It is clear that the smaller of the two has to be chosen,
namely the one where the function is growing.
For this reason we will  call this an ``ascending root cutoff''.
On the other hand if $0<\tilde\mu<\sqrt{27/4}$, equation \eq{rootz} has no positive root
and the ascending root does not exist.
However, we observe that the equation $\tilde\lambda^{T-}(x)=-1$ has a
real positive root for any $\tilde\mu$.
As mentioned in Section 3, in the evaluation of the beta functions
it is not important whether the modes have positive or negative eigenvalue.
Therefore we can use this descending root to define the cutoff.
We call this a ``descending root cutoff''.
When $\tilde\mu<0$, the discussion can be repeated interchanging the roles of $\tilde\lambda^{T+}$ and $\tilde\lambda^{T-}$.

%
\begin{figure}
\begin{center}
{\resizebox{1\columnwidth}{!}
{\includegraphics{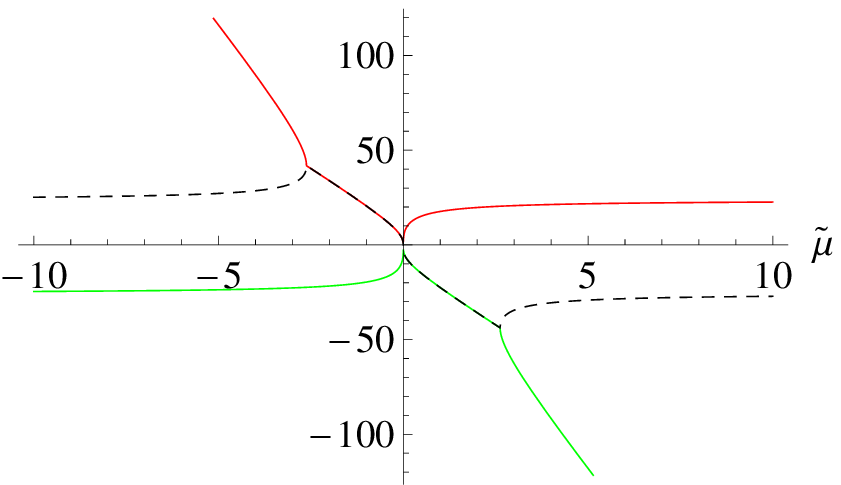}
\includegraphics{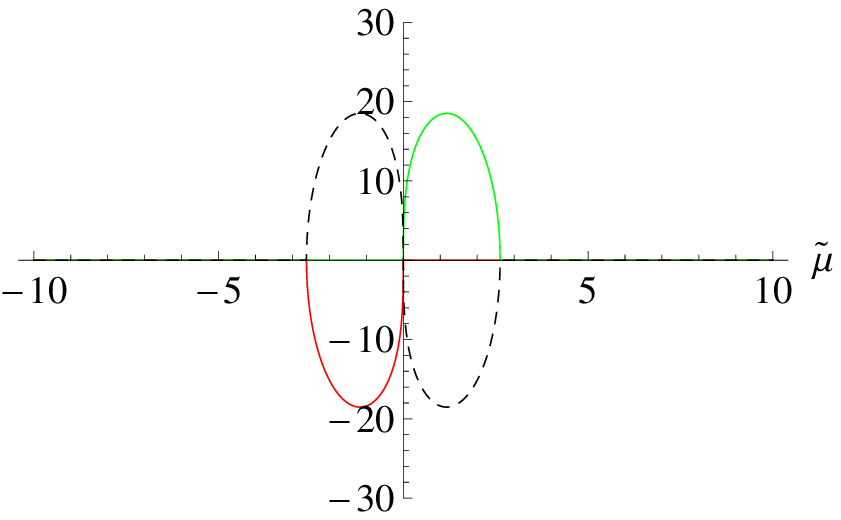}}
\caption{\small
The real (left panel) and imaginary (right panel) parts of the
roots of the equation $\tilde\lambda_n^{T+}=1$, for $\tilde R=\tilde\Lambda=0.01$,
as functions of $\tilde\mu$.
The first root (thick dotted line) always has positive real part and is complex for $-\sqrt{27/4}<\tilde\mu<0$;
the second root (dashed line) always has negative real part and is complex for
$0<\tilde\mu<\sqrt{27/4}$;
the third root (solid line) is complex for $-\sqrt{27/4}<\tilde\mu<\sqrt{27/4}$.
The solutions of the equation $\tilde\lambda_n^{T-}=1$ are obtained by the reflection
$\tilde\mu\to -\tilde\mu$.
}}
\end{center}
\end{figure}
%
%
%
\begin{figure}
\begin{center}
{\resizebox{1\columnwidth}{!}
{\includegraphics{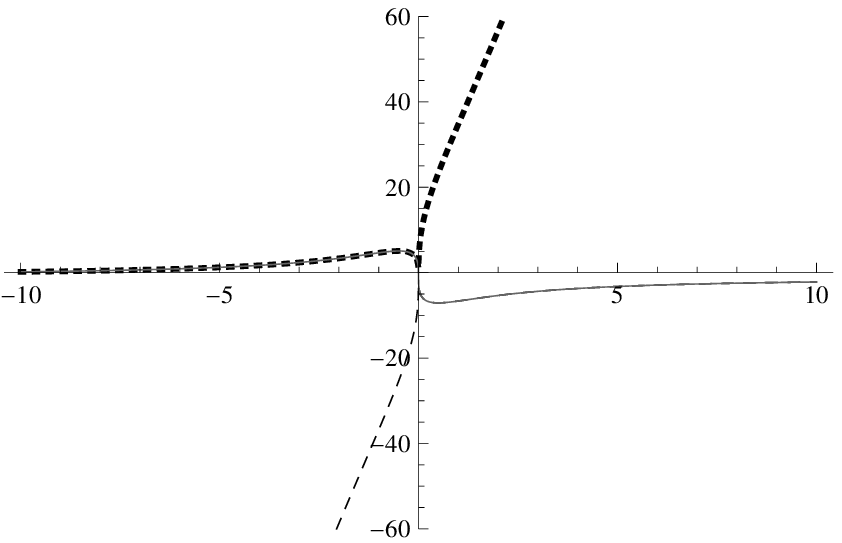}
\includegraphics{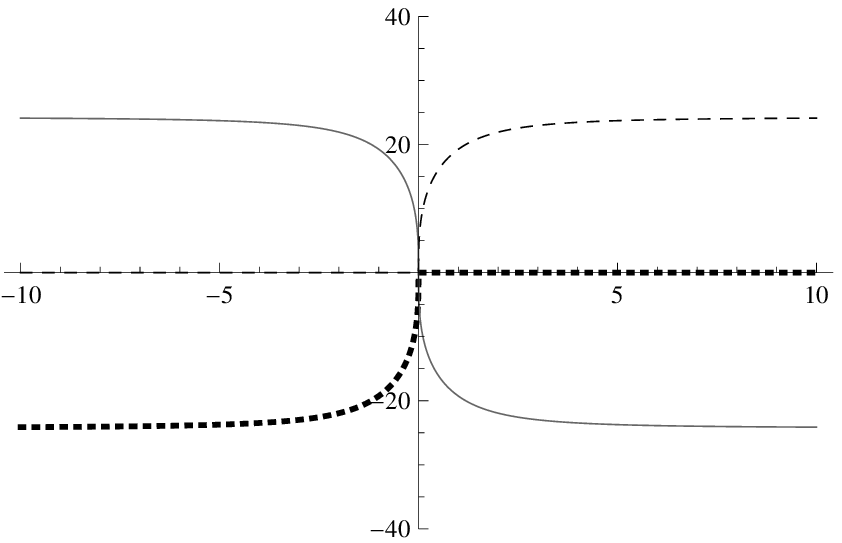}}
\caption{\small The real (left panel) and imaginary (right panel) parts of the
roots of the equation $\tilde\lambda_n^{T-}=-1$, for $\tilde R=\tilde\Lambda=0.01$,
as functions of $\tilde\mu$.
The first root (thick dotted) always has positive real part and is complex for $\tilde\mu<0$;
the second root (dashed) always has negative real part and is complex for
$\tilde\mu>0$;
the third root (solid line) is always complex.
The solutions of the equation $\tilde\lambda_n^{T+}=-1$ are obtained by the reflection
$\tilde\mu\to -\tilde\mu$.
}}
\end{center}
\end{figure}

Since the descending solution of $\tilde\lambda^{T-}=-1$
always exists, while the ascending solution of $\tilde\lambda^{T-}=1$
only esists if $|\tilde\mu|>\sqrt{27/4}$, one wonders why not use always
the former. The reason is that
the information one can get from these different cutoffs is complementary.
For example if we restrict ourselves to $\tilde\mu>0$,
when $\tilde\mu$ becomes very large, the descending solution of
$\tilde\lambda_n^{T-}=-1$ is much larger than the ascending
solution of $\tilde\lambda_n^{T+}=1$.
In fact the sum over negative chirality modes is divergent
in the limit $\tilde\mu\to\infty$. The beta functions computed in this way
will have fictitious singularities in this limit and it will not be possible
to flow smoothly to the case when the CS term is absent.
As we will see later, this also implies that the beta functions
have fictitious singularities near $\tilde G=0$, so it will not be possible
to study the Gaussian FP in this cutoff scheme.
On the other hand, the ascending solution has an imaginary part
when $|\tilde\mu|<\sqrt{27/4}$ and therefore cannot be used to
study the small $\tilde\mu$ region.
The conclusion is that in order to properly understand the behavior of the
theory over the whole range of values of $\tilde\mu$ one has to use
both schemes: the ``descending'' cutoff for small $\tilde\mu$ and
the ``ascending'' cutoff for large $\tilde\mu$.


\section{The Beta Functions}


Applying the Euler-Maclaurin formula to each of the sums in \eq{sums}
and extracting from each of the integrals the appropriate powers of $R$,
the r.h.s. of \eq{sums} can be written
\be
\label{contributions}
\partial_t\C_k = \sum\left[C_0 R^{-3/2}+C_2 R^{-1/2}+C_{3/2}+\frac12 F(n_0)-\frac{B_2}{2!}F'(n_0)\right]
\ee
where the sum is over $h^T$, $\xi$, $\sigma$, $h$, $C^T$, $C$.

The contributions of each spin component to the $c_{3/2}$ term are
independent of the cutoff scheme. They are listed in the following table:

\bigskip

\begin{center}
\begin{tabular}{|c|c|c|c|c|}
\hline
              &$n_0$ & $C_{3/2}$ & $F(n_0)$ & $F'(n_0)$ \\
\hline
$h^T_{\mu\nu}$ & 2 & $12$  & $20$ & $24$ \\
\hline
$\hat\xi^\mu$  & 2 & $-24$  & $32$  & $24$  \\
\hline
$\hat\sigma$   & 2 & $-18$  & $18$  &  $12$ \\
\hline
$h$            & 0 & $-\frac{2}{3}$  &  $2$ &  $4$ \\
\hline
$C^\mu$        & 1 & $-\frac{8}{3}$  & $12$  &  $16$ \\
\hline
$\hat C$       & 1 & $-\frac{16}{3}$  & $8$  & $8$  \\
\hline
\end{tabular}
\end{center}

\bigskip

In the second column, the lower end of integration in \eq{ml} is shown.
Summing all the contributions we conclude that the coefficient of the $R$-independent term is exactly zero.
More precisely, the sum $C_{3/2}+\frac12 F(n_0)-\frac{B_2}{2!}F'(n_0)$
is zero separately for the trace-free part of $h_{\mu\nu}$
(the sum of the first three lines), for the trace part $h$
and for the ghosts (the sum of the last two lines).
These are the components of the fields that can be defined by purely
algebraic conditions. The cancellation does not occur for
components defined by differential constraints, such as the transversality condition.
To some extent, the overall cancellation of the $R$-independent terms is expected.
In the simpler setting of pure gravity without CS term,
or any matter field coupled to gravity,
the sums can be evaluated using the heat kernel expansion.
On a manifold without boundary the trace of the heat kernel
contains only even powers of $R$, and in a 3-dimensional manifold
the volume prefactor is proportional to $R^{-3/2}$, so that
the expansion of $\partial_t\Gamma_k$ contains only odd powers of $R$,
and there is no $R$-independent term.
So the CS term will not be induced, if one starts without it.
In Appendix D we show that the sums done with the Euler-Maclaurin method
explained above exactly reproduce the heat kernel results
for $3D$ gravity without CS term at one loop.
The fact that all the contributions listed in the table above exactly cancel when properly summed is therefore a nontrivial check of our calculation.

To obtain the rest of the beta functional $\partial_t\Gamma_k$ there remain to sum the contributions of type $C_0$ and $C_2$ in \eq{contributions}; the final result has the following structure:
\bea
\partial_t \C_k &=& \frac{V(S^3)}{16\pi}
\Big[k^3A(\tilde\Lambda,\tilde\mu)+ k B(\tilde\Lambda,\tilde\mu) R + O(R^2)\Big]\ ,
\label{mr}
\eea
where we have inserted powers of $k$ such that the $A$- and $B$-coefficients are dimensionless. The volume of $S^3$ with radius $\ell$ is $V(S^3)=2\pi^2 \ell^3$ with $\ell= \sqrt{\frac6{R}}$.

Equation \eq{contributions} is an expansion in $R$, whose coefficients
are functions of $\Lambda$, $\mu$ and $k^2$. For reasons that will become apparent below, we shall restrict ourselves to the parameter region where $\Lambda$ and $R$ are of the same order. Therefore we shall also expand the coefficients in \eq{contributions} in powers of $\Lambda$, namely in $C_0$ we keep at most terms linear in $\Lambda$ while in $C_2$ we only keep the $\Lambda$-independent terms.
We will give explicit expressions for $A$ and $B$ later.

Evaluating the (Euclidean version of the) renormalized TMG action \eq{1} on $S^3$ background, it can be written in the form
\be
\C_k = V(S^3) \left(\frac{ 2\Lambda}{16\pi G}- \frac{1}{16\pi G} R
+\frac{1}{12 \sqrt 6 \pi G \mu} R^{3/2}+ O(R^2)\right)\ ,
\label{gs}
\ee
where we have used that the integral of the  CS term on $S^3$ is given by $\int \tr (\omega d\omega +\frac23 \omega^3)=32 \pi^2$.
The couplings $\Lambda$, $G$, $\mu$ are now renormalized couplings evaluated at scale $k$.
Rescaling the coupling constants as
\be
G = \tilde G k^{-1}\ ,\qquad \Lambda=\tilde\Lambda k^2\ ,\qquad
\mu= \tilde\mu k\ ,
\ee
so as to make them dimensionless, and comparing the $t$-derivative of \eq{gs} with \eq{mr}, we obtain:
\bea
\frac{1}{8\pi \tG}\left(\partial_t \tL
-\frac{\partial_t\tilde G}{\tilde G}\tL\right) &=&
- \frac{3\tL}{8\pi \tG}+\frac{A}{16\pi}\ ,
\w2
\frac{\partial_t\tilde G}{16\pi \tG^2}  &=&\frac{1}{16\pi \tG}
+\frac{B}{16\pi}\ ,
\w2
\frac{1}{12\sqrt{6}\pi\tilde\mu\tG}
\left(\frac{\partial_t\tilde G}{\tilde G}+\frac{\partial_t\tilde \mu}{\tilde \mu}\right) &=& 0\ ,
\label{beta}
\eea
The last equation results from the fact that the terms of order $R^{3/2}$ in \eq{mr} cancel, and it implies that the dimensionless combination $\nu \equiv G\mu=\tilde G\tilde\mu$ does not run. From the other two equations one obtains the one loop beta functions of $\tilde G$ and $\tilde\Lambda$:
\bea
\partial_t\tilde G &=&\tilde G+B(\tilde\mu)\tilde G^2\ ,
\nonumber\w2
\partial_t\tilde\Lambda &=& -2\tilde\Lambda
+\frac{1}{2}\tilde G\left(A(\tilde\mu,\tilde\Lambda)+2B(\tilde\mu)\tilde\Lambda\right)\ .
\label{oneloopbeta}
\eea
These equations have exactly the same form as in pure gravity with
cosmological constant, except that the coefficients $A$ and $B$
are now $\tilde\mu$-dependent.
In order to solve for the RG flow, we use that $\nu$ does not run and substitute $\tilde\mu = \nu/\tilde G$. This yields two ordinary first order differential equations for $\tilde G(t)$ and $\tilde\Lambda(t)$, depending on the fixed constant value of the external parameter $\nu$.
Rather surprisingly, despite the very different functional form, the resulting flow is numerically quite similar to that of pure gravity with cosmological constant. Unlike in pure gravity, however, here we cannot give the solution in closed form.
We will now describe the results for different cutoff schemes.


\section{Results}


\subsection{Ascending root cutoff}


For $|\tilde\mu|>\sqrt{27/4}$ we define the cut on the spin two modes
as the smallest root of the equation $\tilde\lambda^{T\pm}=1$.
In the case of positive $\tilde\mu$,
the cutoff on $\tilde\lambda^{T+}$ corresponds
to the thick dotted line in Fig 2, while the cutoff on $\tilde\lambda^{T-}$
is obtained by reflecting the solid line.
(Note that the two roots are different, so that the number of modes
of positive and negative chirality is different.)
Calculating the sums yields beta functions that are real and well defined for
$0<\tilde G<\sqrt{4/27}\nu$.
For negative $\tilde\mu$ the roles of $\tilde\lambda^{T+}$ and
$\tilde\lambda^{T-}$ are interchanged.
Calculating the sums yields beta functions that are real and well defined for
$-\sqrt{4/27}\nu<\tilde G<0$.
The two calculations match smoothly along the line $\tilde G=0$,
so one can put them together to obtain a RG flow
on the whole region $\tilde G^2<4\nu/27$.

With this prescription we calculate the coefficients $A$ and $B$.
These arise as complicated functions involving cubic roots,
but after some manipulations can be reduced to the following relatively
simple form:
\bea
A(\tilde\Lambda,\tilde\mu) &=&
-\frac{16}{3\pi}
+\frac{9(2\sqrt{3}\cos2\theta-\sqrt{3}\cos4\theta
+8(\cos\theta)^3\sin\theta)}
{\pi(\cos 3\theta)^3}
\w2
&&
+\frac{8(3+11\alpha-2\alpha^2)}{\pi(4-\alpha)}\tilde\Lambda
+\frac{48(\cos\theta-\sqrt{3}\sin\theta)}{\pi\sin 6\theta}\tilde\Lambda\ ,
\nn\w4
B(\tilde\mu)&=&-\frac{4(1+\alpha)(11-2\alpha)}{\pi(4-\alpha)}
-\frac{2(\sqrt{3}\sin\theta-\cos\theta) +22(\sqrt{3}\sin 5\theta+\cos 5\theta))}{3\pi\sin 6\theta }\ ,\nn
\eea
where we have introduced the angle
\be
\theta=\frac{1}{3} \arctan \sqrt{\frac{4\tilde\mu^2}{27} -1}\ .
\label{theta}
\ee

The beta functions admit a Taylor expansion around $\tilde\Lambda=\tilde G=0$
\bea
\label{expandasc}
\partial_t\tilde G&=&
{\tilde G}-\frac{2(94-4\alpha^2)\tilde G^2}{3\pi(4-\alpha)}
-\frac{95{\tilde G}^4}{6\pi\nu ^2}
-\frac{2233 {\tilde G}^6}{32\pi\nu ^4}
+O\left({\tilde G}^7\right)\ ,
\w2
\partial_t\tilde\Lambda &=&-2\tilde\Lambda
-\frac{4(14-27\alpha+4\alpha^2)\tilde G\tilde\Lambda}{3\pi(4-\alpha)}
+\frac{\tilde G^3(42+115\tilde\Lambda)}{6\pi\nu^2}
+\frac{11 {\tilde G}^5 (78+343\tilde\Lambda)}{32\pi\nu^4}
+O\left({\tilde G}^6\right)\ .
\nn
\eea

We observe, as a nontrivial check, that
in the limit $\tilde\mu\to\infty$
the beta functions agree with the result for gravity without CS term,
which are calculated in Appendix D using a different method.

%
\begin{figure}
\begin{center}
{\resizebox{1\columnwidth}{!}
{\includegraphics{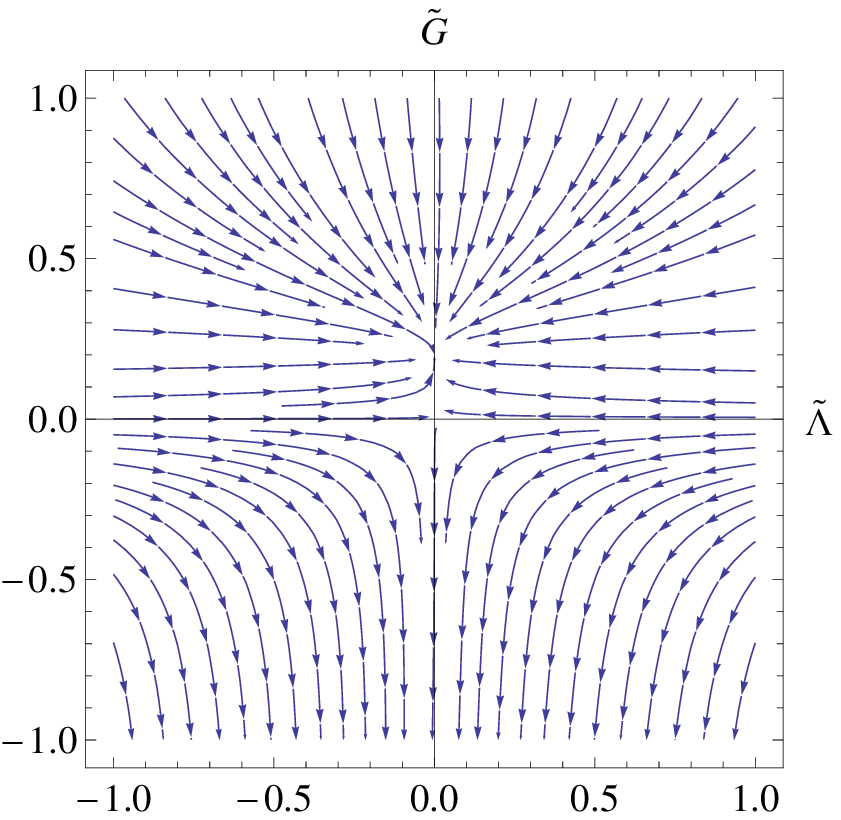}
\includegraphics{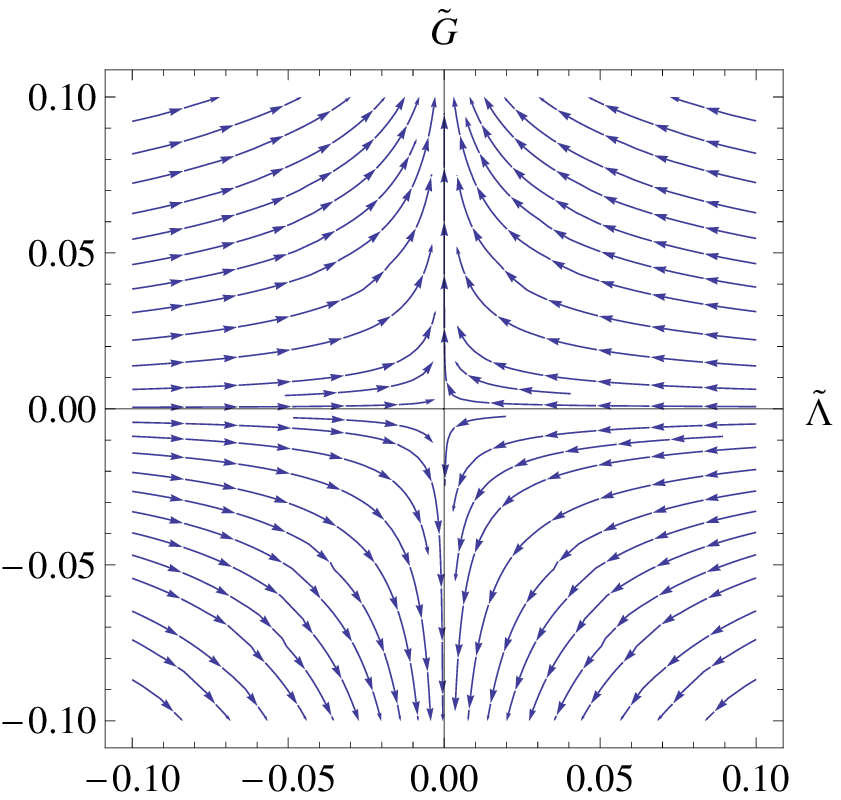}}
\caption{\small The flow in the $\tilde\Lambda$-$\tilde G$ plane
for $\alpha=0$, $\nu=5$, in the ascending root cutoff scheme.
Right panel: enlargement of the region around the origin,
showing the Gaussian FP.
The beta functions become singular at $|\tilde G|=1.9245$,
outside the domain of the picture, but this singularity is
an artifact of the scheme.
}}
\end{center}
\end{figure}

The flow is shown, in the case $\nu=5$, in Figure 4.
We note the main features, which hold for any value of $\nu>0$.
There is a FP, called the Gaussian FP, in the origin,
which is seen to be attractive in the $\tilde\Lambda$ direction and
repulsive in the $\tilde G$ direction.
To make this statement more quantitative, one has to study the
linearized equation
\be
k\frac{d(\tilde g_i - \tilde g_{i\star})}{dk}= M_{ij} (\tilde g_j - \tilde g_{j\star})\ .
\ee
where
\be
M_{ij} = \frac{\partial\beta_i}{\partial{\tilde g}_j}{\Big |}_\star \ ,
\qquad
\beta_i = k\frac{d{
\tilde g}_i}{dk}\ ,
\ee
Here $\tilde g_i$ are the couplings and $\tilde g_{i*}$ their values at the FP.
At the Gaussian FP $\tilde g_{i*}=0$ one finds
\be
M_{ij} =
\left(
  \begin{array}{cc}
    -2 &  0  \\
    0 & 1 \\
  \end{array}
\right)
\ee
The eigenvalues of this matrix (``scaling exponents'')
correspond to the canonical mass dimensions of $\Lambda$ and $G$,
as expected. An important fact (which had been previously observed in the
absence of CS term in \cite{Lauscher}) is that the eigenvectors coincide with the
coordinate axes. This is not the case in higher dimensional gravity,
where the eigenvector with eigenvalue equal to the canonical dimension of $G$ has a novanishing component along $\tilde\Lambda$.

In addition to the Gaussian FP there is a nontrivial FP at
$\tilde \Lambda_*=0.000490471$ and $\tilde G_*=0.200016$.
It is seen to be UV attractive in both directions.
The eigenvalues of the corresponding stability matrix are found to
be equal to $-2.29401$ and $-1.00515$.

The position of the FP and the eigenvalues of the stability matrix
are given, for other values of $\nu$, in Figure 5.
Note that $\tilde \Lambda$ is always positive but very small.
This is due to the absence of a term of order $\tilde G^2$
in the expansion of $\partial_t\tilde\Lambda$ in \eq{expandasc}.
Actually if one plots the contours of the beta functions for $\tilde\Lambda$
one may see this as an effect of the deformation of the flow
due to the presence of the boundary at $\tilde G=\nu\sqrt{4/27}$
(which occurs at $\tilde G=1.924$ in Figure 4).
This can probably be regarded as a scheme artifact.

%
\begin{figure}
\begin{center}
{\resizebox{1\columnwidth}{!}
{\includegraphics{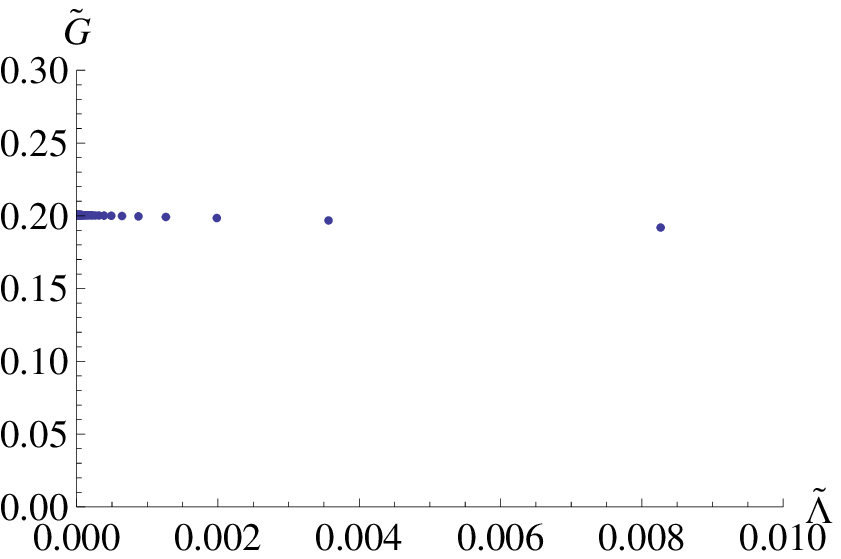}
\includegraphics{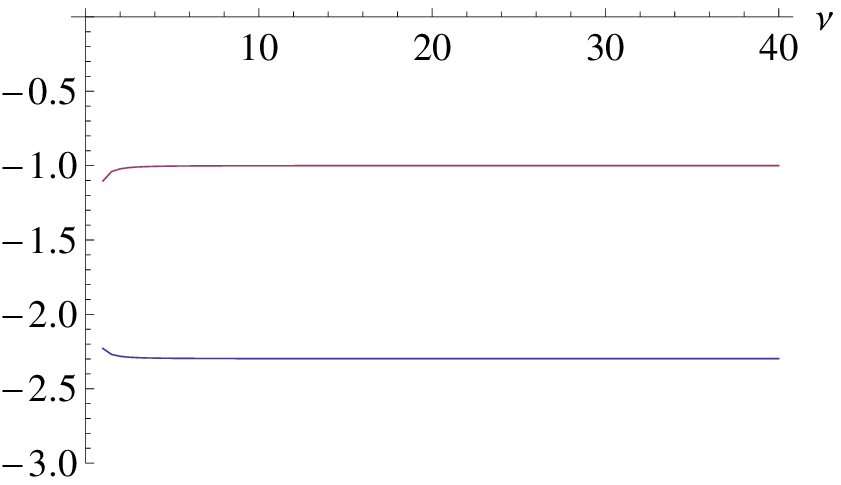}}
\caption{\small Position of the FP (left panel)
and eigenvalues of the stability matrix (right panel)
for the nontrivial FP with $\alpha=0$, $1<\nu<40$, in the ascending root cutoff scheme.
Note that for this range of $\nu$ the singularity is always above the FP.
In the left panel, $\nu$ grows from right to left. Note that $\tilde\Lambda_*>0$
in this scheme. See Section 7.1 for a discussion.
For large $\nu$, $\tilde G_*$ tends to $0.2005$ and the eigenvalues tend to $-1$ and $-2.298$.
}}
\end{center}
\end{figure}

Finally we observe that in the limit $\tilde\mu\to\infty$ the beta functions
reduce to
\bea
\partial_t\tilde G &=&
\tilde G-\frac{4}{3\pi}\,\frac{47-2\alpha^2}{4-\alpha}\tilde G^2\ ,
\nonumber\w2
\partial_t\tilde\Lambda &=& -2\tilde\Lambda
-\frac{4}{3\pi}\frac{14-27\alpha+4\alpha^2}{4-\alpha}\tilde\Lambda\tilde G\ .
\label{oneloopbetaasc}
\eea
which agree with the result for pure gravity without CS term.


\subsection{Descending root cutoff}


In this scheme the cut on the positive and negative chirality
spin two modes is defined by two different equations,
as follows:

\smallskip
\begin{center}
	{\centering
		\begin{tabular}
			{|c|c|}\hline
 $\tilde\mu>0$ & $\tilde\mu<0$ \\\hline
$\tilde\lambda^{T+}=1$ & $\tilde\lambda^{T+}=-1$  \\
$\tilde\lambda^{T-}=-1$ & $\tilde\lambda^{T-}=1$ \\\hline
		\end{tabular}
		}		
\end{center}

\smallskip
\noindent
Using the criteria described in the previous section,
in the case of positive $\tilde\mu$,
the cutoff on $\tilde\lambda^{T+}$ corresponds once again
to the thick dotted line in Fig 2, while the cutoff on $\tilde\lambda^{T-}$
is the thick dotted line in Fig 3.
For negative $\tilde\mu$ the roles of $\tilde\lambda^{T+}$ and
$\tilde\lambda^{T-}$ are interchanged.
Again the two roots are different, so that the number of modes
of positive and negative chirality is different.
In fact in this case the behavior is drastically different,
since the cut on the descending mode grows linearly with $\tilde\mu$.
This will give a singularity in the beta functions for $\tilde\mu\to\infty$,
and therefore, when we make the substitution $\tilde\mu=\nu/\tilde G$,
for $\tilde G\to 0$. Thus this scheme is really useful only for
sufficiently small $\tilde\mu$.

With this prescription we find:
\bea
A (\tm, \tilde\Lambda) &=& -\frac{16}{3\pi} + \frac{\sqrt{3}}{\pi} \left[
\left(\frac{2\cosh2\eta-1}{\cosh3\eta}\right)^3
+
\left(\frac{2\cosh2\psi+1}{\cosh3\psi}\right)^3\right]
\w2
&& \!\!\!\!\!\!\!
+\frac{8(3+11\alpha-2\alpha^2)}{\pi(4-\alpha)} \tilde\Lambda
+ \frac{8\sqrt{3}}{\pi}\left[
\frac{1}{2\cosh\eta+\cosh3\eta}
+\frac{1}{2\sinh\psi-\sinh3\psi}
\right]\tilde\Lambda
\nn\w4
B (\tm) &=& -\frac{4(11+9\alpha-2\alpha^2)}{3\pi(4-\alpha)}
-20\sqrt{3}\frac{\cosh2\eta+\cosh2\psi}{3\pi \cosh3\eta}
\nn\w2
&&
+\frac{2}{3\sqrt{3}\pi} \left(\frac{8\cosh2\eta-1}{\cosh3\eta+2\cosh\eta} +\frac{8\cosh2\psi+1}{\sinh3\psi-2\sinh\psi}\right)
\nn
\eea
where we have defined
\be
\eta = \frac13\, \mathrm{arctanh}\, \sqrt{1-\frac{4\tm^2}{27}}\ , \qquad
\psi = \frac13\, \mathrm{arccoth}\, \sqrt{1+\frac{4\tm^2}{27}}\ .
\label{eta}
\ee
%
%
\begin{figure}
\begin{center}
{\resizebox{1\columnwidth}{!}
{\includegraphics{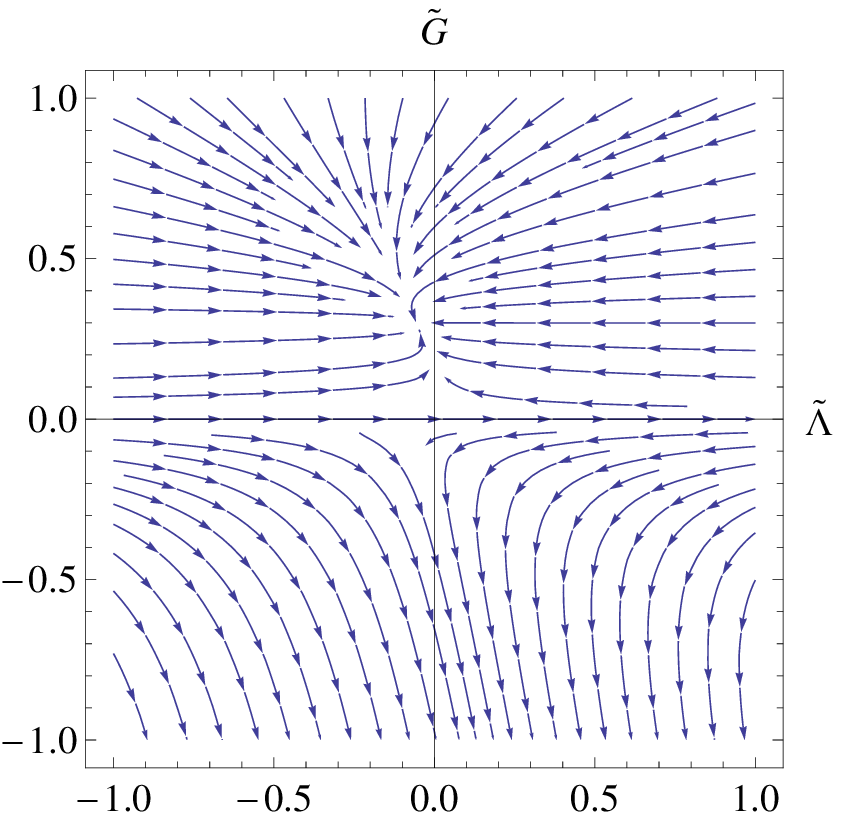}
\includegraphics{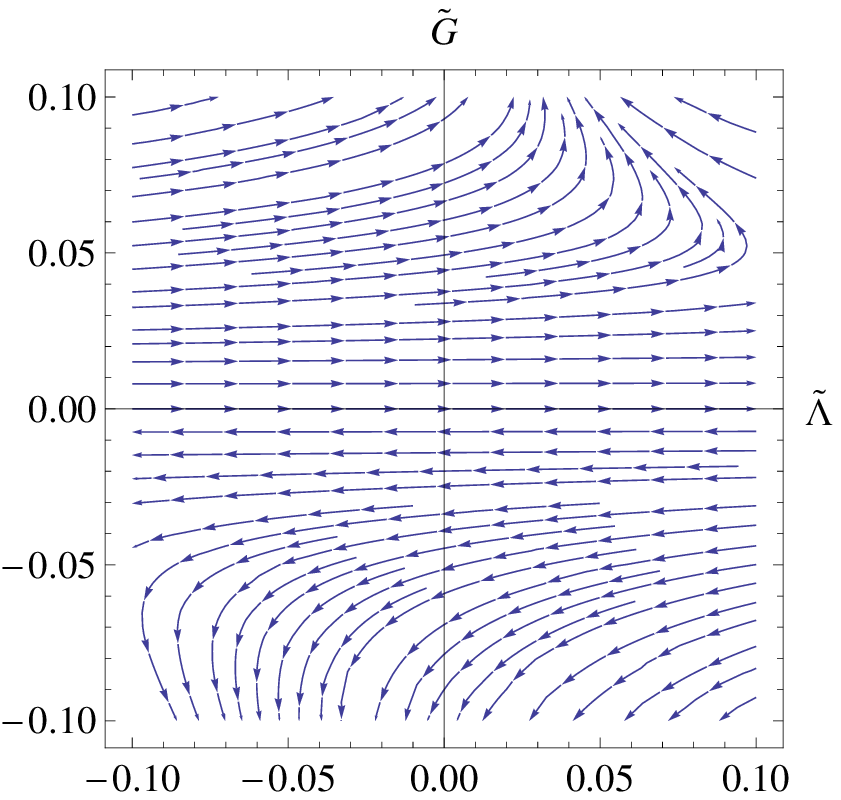}}
\caption{\small The flow in the $\tilde\Lambda$-$\tilde G$ plane
for $\alpha=0$, $\nu=0.1$, in the descending root cutoff scheme.
Right panel: enlargement of the region around the origin,
showing that there is no Gaussian FP.
The beta functions diverge on the $\tilde\Lambda$ axis.
}}
\end{center}
\end{figure}
\noindent

A representative flow is shown in Figure 6 for $\nu=0.1$.
Superficially this may look similar to Figure 4, but there are
some important differences.
First and foremost, there is no Gaussian FP, as is clear from
the enlargement of the area around $\tilde\Lambda=\tilde G=0$.
The FP is wiped out by the singularity at $\tilde G=0$,
where the beta function of $\tilde\Lambda$ blows up.
Note that the flow lines run in opposite directions on the two
sides of this line.
All the flow lines which, for positive $\tilde G$,
seem to come in from plus infinity actually originate from minus infinity,
and follow closely the $\tilde G$ axis before bending over.
Similarly for $\tilde G<0$ all the flow lines that seem to come in from
minus infinity actually come from plus infinity and run close to the
$\tilde G=0$ line.
There is thus an infinite accumulation of flow lines along the $\tilde\Lambda$ axis.
Another difference with Figure 4 is that the flow lines near the $\tilde G$ axis
are tilted in the opposite direction, and by a much larger amount.

For $\nu=0.1$ the nontrivial FP occurs at $\tilde\Lambda_*=-0.0565337$
$\tilde G_*=0.30036$ and the eigenvalues of the stability matrix are
$-2.76746$, $-0.781453$.
The position of the FP and the eigenvalues, for $10^{-6}<\nu<0.5$
are shown in Figure 7.
The FP occurs at positive $\tilde\Lambda$ for $\nu>0.18$ and negative
$\tilde\Lambda$ for $\nu<0.18$.
Given that this cutoff scheme is more dependable for small $\tilde\mu$
(and hence, at fixed $\tilde G$, for small $\nu$), it is again possible
that the positive values of $\tilde\Lambda$ are a scheme artifact.

%
\begin{figure}
\begin{center}
{\resizebox{1\columnwidth}{!}
{\includegraphics{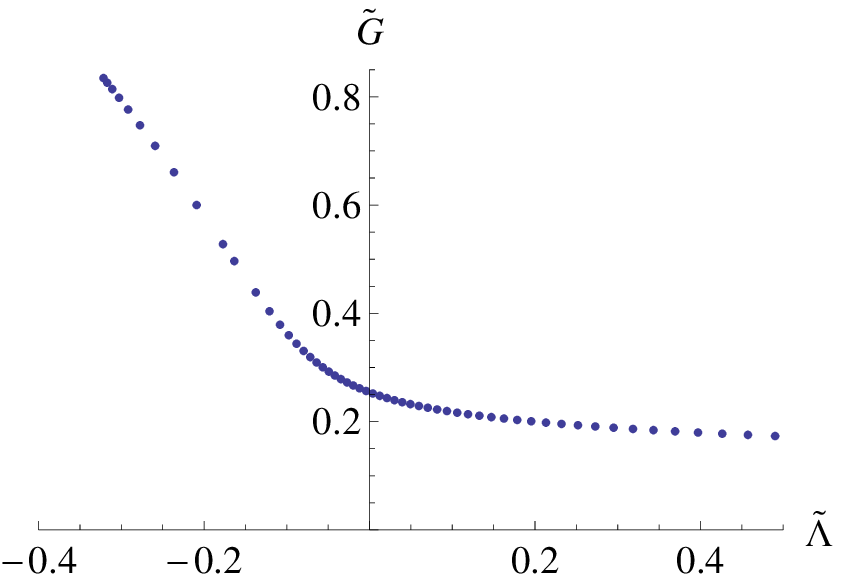}
\includegraphics{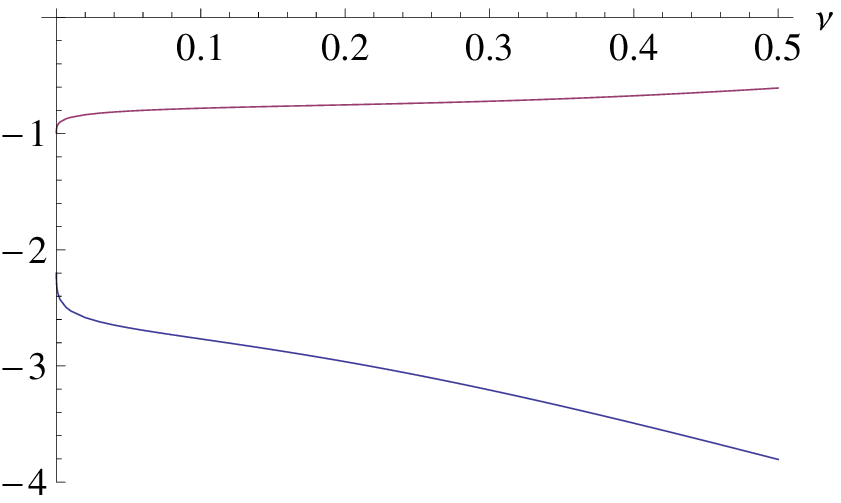}}
\caption{\small Position of the FP (left panel)
and eigenvalues of the stability matrix (right panel)
for the nontrivial FP with $\alpha=0$, $10^{-6}<\nu<0.5$,
in the descending root cutoff scheme.
In the left panel, $\nu$ decreases from right to left.
The cosmological constant changes sign for $\nu=0.18$.
The rightmost point ($\nu=0.5$) has $\tilde\mu\approx 3>\sqrt{27/4}$
and therefore is in the region where the scheme becomes unreliable.
}}
\end{center}
\end{figure}
Just as the beta functions in the ascending root scheme
tend to a very simple form in the limit $\tilde\mu\to\infty$,
the beta functions in the descending root scheme
tend to a very simple form in the limit $\tilde\mu\to 0$:
\bea
\partial_t\tilde G &=&
\tilde G-\frac{4}{3\pi}\,\frac{11+9\alpha-2\alpha^2}{4-\alpha}\tilde G^2\ ,
\nonumber\w2
\partial_t\tilde\Lambda &=& -2\tilde\Lambda
-\frac{8}{3\pi}\left(1+\frac{1-12\alpha+2\alpha^2}{4-\alpha}\tilde\Lambda\right)\tilde G\ .
\label{oneloopbetadesc}
\eea
These beta functions have a FP at $\tilde\Lambda_*=-1/3$,
$\tilde G_*=\frac{3\pi(4-\alpha)}{4(11+9\alpha-2\alpha^2)}$,
which for $\alpha=0$ is $\tilde G_*\approx 0.8568$.
The eigenvalues of the stability matrix are $-2.182$ and $-1$
corresponding to the directions $(1,0)$ and $(-0.5499, 0.8352)$.

\subsection{Spectrally balanced cutoff}

In the preceding sections we have described two calculations
of the beta functions for TMG.
Both schemes are ``spectrally unbalanced'' in the sense that the
summations over the spin two fields in \eq{sums} contained a different number of
positive and negative chirality modes.
Consideration of the behavior of the roots in Figures 1, 2 and 3 show that
the ascending scheme becomes balanced for large $\tilde\mu$ while
the descending scheme becomes balanced for small $\tilde\mu$.
These are the regimes where these schemes are most reliable.
One thus wonders whether one can work in a scheme where the sums always
contain equal numbers of positive and negative chirality modes, by construction.

Such a scheme can be constructed by tweaking the cutoff profile.
First, we must allow different profiles for different spin components.
The spin one and zero components will be treated as before.
For the spin two components we allow
the more general form $R_k(z)=(q(z)k^2-z)\theta(q(z)k^2-z)$
where $q(z)$ is a dimensionless function.
For any choice of the function $q(z)$ we have
$\partial_t R_k(z)=2k^2\theta(q(z)k^2-z)$ and for $z<k^2$, $P_k(z)=k^2$,
so that $W(z)=2\theta(q(z)-z/k^2)$.
With this cutoff choice the traces are still finite sums of the multiplicity,
up to a maximal value $n_{\rm max}$ which is determined as a root of the equation
\be
\tilde\lambda(x)=q(\tilde\lambda(x))\ .
\ee
Thus the sums are still calculable by the same technique used previously.
A choice of $n_{\rm max}$ implicitly defines a choice of the function $q$.
The sums on the spin two components in \eq{sums} are now replaced by
\be
\sum_{\pm}\sum_{n=2}^\infty m_n^{T\pm}\theta(q^\pm(\tilde\lambda_n^{T\pm})-\tilde\lambda_n^{T\pm})=
2\sum_{n=2}^{n_{\rm max}}m_n^T\ .
\ee
If $\tilde\mu>0$, we choose $n_{\rm max}$ to be the unique positive root of the equation $\tilde\lambda_n^{T+}=1$; then we cut both sums at this value.
In this case the function $q$ is $q^+=1$ for the positive chirality modes,
while $q^-$ is determined by the condition
$q^-(\tilde\lambda_n^{T-}(x))=\tilde\lambda_n^{T+}(x)/\tilde\lambda_n^{T-}(x)$.
Likewise, if $\tilde\mu<0$ we cut at the root of $\tilde\lambda_n^{T-}=1$.

With this prescription we get the following results:
\bea
A(\tilde\mu,\tilde\Lambda) &=&
-\frac{16}{3\pi}
+\frac{2\sqrt{3}}{\pi (\cosh\eta)^3}
\label{ab}\w2
 &&
 +\frac{8(3+11\alpha-2\alpha^2)}{\pi(4-\alpha)}\tilde\Lambda
+  \frac{16\sqrt{3}}{\pi(\cosh 3\eta+2\cosh\eta)}\tilde\Lambda \ ,
\nn\w4
B(\tilde\mu) &=&
-\frac{4(11+9\alpha-2\alpha^2)}{3\pi(4-\alpha)}
-\frac{8\sqrt{3}}{9\pi} \left(
\frac{8+11 \cosh 2\eta}{\cosh 3\eta+2\cosh\eta}\right)\ ,
\nn
\eea
where $\eta$ is given by \eq{eta}.  Thus, we observe that the apparent poles at $a=0$, which correspond to $\eta=0$, are actually absent as can be readily deduced from \eq{ab}, and the coefficients $A$, $B$ are real. Note also that $\eta=i\theta$ where $\theta$ is the angle that we have encountered in Section 7.1 (see \eq{theta}. Furthermore $\cosh 3\eta= 3\sqrt{3}/|\tm|$, and the results depend on the absolute value of $\tm$.


In the limit $\tilde\mu \rightarrow \infty$ we get
\be
A\to \frac{8(11+9\alpha-2\alpha^2)}{\pi(4-\alpha)} \tilde\Lambda\ ,\qquad
B\to -\frac{4(47-2\alpha^2)}{3(4-\alpha)}\ .
\label{infty}
\ee
This limit corresponds to neglecting the CS term, and it is reassuring that we find agreement with the result of the heat kernel calculation in Appendix D.

The Taylor expansion of the beta functions in powers of $\tilde G$
around the Gaussian FP is
\bea
\label{expand}
\partial_t\tilde G&=& {\tilde G}-\frac{4(47-2\alpha^2)\tilde G^2}{3\pi(4-\alpha)}+\frac{28{\tilde G}^3}{3\pi\nu}
-\frac{95{\tilde G}^4}{6\pi\nu ^2}
+O\left({\tilde G}^5\right)\ ,
\w2
\partial_t\tilde\Lambda &=&-2\tilde\Lambda
-\frac{4(14-27\alpha+4\alpha^2)\tilde G\tilde\Lambda}{3\pi(4-\alpha)}
-\frac{4{\tilde G}^2(3+5\tilde\Lambda)}{3\pi\nu}
+\frac{\tilde G^3(42+115\tilde\Lambda)}{6\pi\nu^2}
+O\left({\tilde G}^4\right)\ .
\nn
\eea
The $\nu$-independent terms coincide with the result for pure gravity,
as described in Appendix D.

In the limit $\tilde\mu\to 0$ the beta function of $\tilde\Lambda$
has the same expression as in \eq{oneloopbetadesc},
but the beta function of $\tilde G$ becomes singular.
Still, the position of the FP seems to approximate closely the one that
was found in the descending root cutoff. For $\nu=10^{-6}$ we find
$\tilde\lambda_*=-0.315$, $\tilde G_*=0.830$.
This should not come as a surprise, since in this limit the
two positive roots of the equations $\tilde\lambda^{T+}=1$
and $\tilde\lambda^{T-}=-1$ become equal, so the
descending root cutoff becomes spectrally balanced.

%
%
%
\begin{figure}
\begin{center}
{\resizebox{1\columnwidth}{!}
{\includegraphics{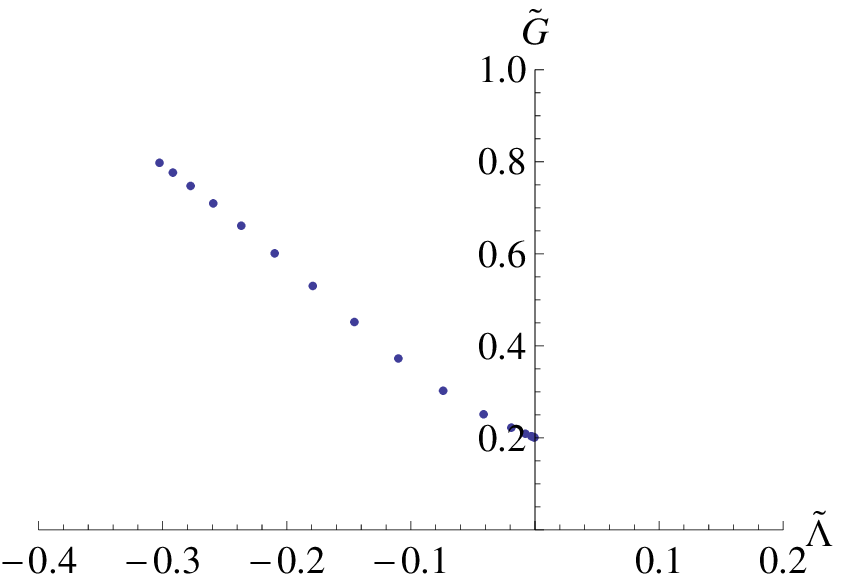}
\includegraphics{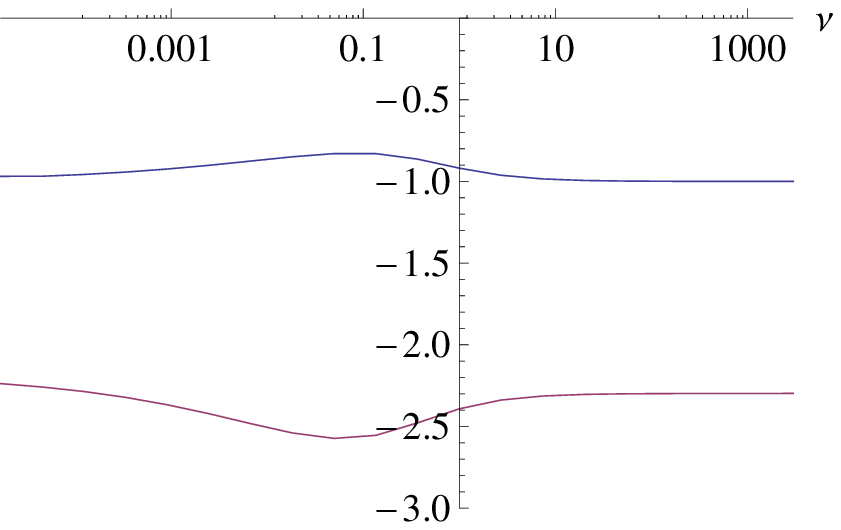}}
\caption{\small Left panel: The position of the FP in the $\tilde\Lambda$-$\tilde G$ plane,
varying $\nu$ from 0.002 (upper left) to 1000 (lower right).
The point with coordinates (0,0.2005) is the limit $\nu\to\infty$.
Right panel: The eigenvalues of $M$ as functions of $\nu$.
For $\nu=0.002$ they are $-1$ and $-2.298$ while for
$\nu=1000$ they are $-0.969$ and $-2.238$.
}}
\end{center}
\end{figure}
%
%


\section{Discussion}


\subsection{Summary}

We begin by summarizing the main results.
We have calculated the beta functions of $\Lambda$, $G$ and $\mu$
and we have studied the corresponding RG flows.
It is generally the case that the beta functions of dimensionful couplings
(by which we mean also the beta functions of the corresponding dimensionless
ratios that we denoted by a tilde) are scheme dependent even to leading order.
This is in contrast to the more familiar case of the dimensionless couplings,
whose beta functions are scheme independent to leading order.
It is therefore not too surprising that the three schemes we have considered
give different pictures.
Out of these different pictures one can however extract some common features
that are presumably scheme-independent.
We summarize them here.

The first main result is that the dimensionless combination $\nu=\mu G$,
which is the coefficient of the CS term, does not run.
We will comment further on this in the subsection on topological properties.
Due to this fact, the remaining beta functions describe a flow in the
$\tilde\Lambda$-$\tilde G$ plane, whose properties depend on the value
of the fixed constant $\nu$.
This two dimensional flow is governed by two FPs:
a Gaussian FP in the origin and a non-Gaussian FP at positive $\tilde G$.
The Gaussian FP is UV attractive in the $\tilde\Lambda$ direction and
UV repulsive in the $\tilde G$ direction, with scaling exponents equal to
the canonical dimensions of $\Lambda$ and $G$.
This FP is not seen in the descending root cutoff scheme, but this is
clearly an artifact of the unphysical singularity of the flow for large $\mu$
(hence small $\tilde G$).
For the same reason one should not trust the properties of the
Gaussian FP that reappears in that scheme in the limit $\nu\to 0$.

There is then a nontrivial FP which always occurs for positive $\tilde G$.
Whether this FP has positive or negative cosmological constant seems to be
somewhat unclear.
The ascending root cutoff, which is reliable for large $\nu$,
and the spectrally balanced cutoff both indicate that for large $\nu$
the FP tends towards the FP that is known to occur in the absence of CS term,
which has $\tilde\Lambda_*=0$.
The descending root cutoff, which is reliable for small $\nu$,
and the spectrally balanced cutoff both indicate that for small $\nu$
the FP tends towards the values $\tilde\Lambda_*\approx -0.33$, $\tilde G_*\approx 0.85$.
We conclude that for small $\nu$ the cosmological constant is certainly negative.
The fact that for $\nu>0.18$ in the descending cutoff scheme $\tilde\Lambda_*$
becomes positive is probably an unphysical effect of the singularity occurring
at $\tilde G=0$, as Figure 7 suggests.
Whether for intermediate values of $\nu$ $\tilde\Lambda_*$ could be positive
is an issue that we leave for future more refined analyses.
On balance, we think that Figure 8 probably gives the most reliable picture.

The final universal feature of the flow is that the nontrivial FP is
UV attractive in both directions, and the existence of a trajectory that
connects the nontrivial FP in the UV to the Gaussian FP in the IR.
The scaling exponents of the nontrivial FP depend on $\nu$
but both for very large and very small $\nu$ they tend to similar values
which are close to -1 and -2.3 (exactly -1, for large $\nu$).

\subsection{Small $\tilde G$ expansion}

We have confined ourselves to the one loop approximation,
which is the lowest order in an expansion in $\tilde G$.
From this point of view one could be surprised by the fact that the beta functions seem to contain arbitrarily high powers of $\tilde G$
(see equations \eq{expandasc}, \eq{expand}).
But note that the original beta functions of $\tilde\Lambda$, $\tilde G$ and $\tilde\mu$
are indeed at most quadratic in $\tilde G$:
It is only when we use the constancy of $\nu$ to solve for $\tilde\mu$ as a function of
$\tilde G$ that the beta functions become nonpolynomial in $\tilde G$.
Since the beta functions will anyway receive higher order corrections,
one may want to truncate the expansion in \eq{expand} to terms at most
quadratic in $\tilde G$.

In the case of the ascending root cutoff there are no terms of order $\tilde G^2$
beyond those that are already present in the absence of the CS term,
so to leading order in that scheme the FP is the same as in the absence of CS term.
In the case of the descending root cutoff one cannot even pose the question,
since the beta functions are singular in the origin.
In the case of the spectrally balanced cutoff one obtains
\bea
\partial_t\tilde G&=& {\tilde G}-\frac{4(47-2\alpha^2)\tilde G^2}{3\pi(4-\alpha)}\ ,
\w2
\partial_t\tilde\Lambda &=&-2\tilde\Lambda
-\frac{4(14-27\alpha+4\alpha^2)\tilde G\tilde\Lambda}{3\pi(4-\alpha)}
-\frac{4{\tilde G}^2(3+5\tilde\Lambda)}{3\pi\nu}\ .
\nn
\eea
In this case the beta function for $\tilde G$ is the same as for pure gravity
without CS term, but the beta function of $\tilde\Lambda$ receives a $\nu$-dependent correction. In this approximation the location of the FP is at
\be
\tilde G_*=\frac{3\pi}{47}\ ;\qquad
\tilde\Lambda_*=-\frac{3\pi}{5\pi+423\nu}\ .
\ee
and the eigenvalues of the linearized flow equation are $-1$ and
$-\frac{108}{47}-\frac{60\pi}{2209\nu}$.

In any case we believe that the vanishing of the beta function of $\nu$ will be true also when higher order corrections are taken into account. This provides some justification for keeping also the higher order terms in the beta functions.

\subsection{The chiral point}

One of the motivations for this work was to determine whether the
chiral point defined by the condition $\mu=\pm\sqrt{|\Lambda|}$
has any special properties under RG.
We can recast this issue in a slightly more general way as follows.
With the dimensionful couplings $\Lambda$, $G$ and $\mu$ one can
construct the dimensionless combinations \eq{dimless}.
The more general question is whether any of these combinations
is invariant under the flow.
We have seen that this is the case for $\nu$.
The conditions $\phi=const$, $\tau=const$ determine two dimensional
subspaces in the space of couplings.
The intersection of these subspaces with the $\tilde\Lambda$-$\tilde G$ plane
defined congruences of curves.
The lines of constant $\tau$ are hyperbolas with equation
\be
\tilde G^2=\frac{\tau}{\tilde\Lambda}
\ee
Since $\nu$ is RG invariant, using to the relation
\be
\label{relation}
\phi=\frac{\nu}{\sqrt{\tau}}
\ee
we see that also $\phi$ is constant on these hyperbolas.
It is easy to see that the flow lines will generally not
preserve $\phi$: they will intersect the lines of constant $\phi$.
If we choose the initial conditions for the flow to lie on one hyperbola,
then the flow will move away from it.
In this sense, the condition $\phi=1$ is not a RG invariant.

One can still ask whether there exist some FP that satisfies
the chirality condition.
For each value of $\nu$, the corresponding FP in the $\tilde\Lambda$-$\tilde G$ plane determines also a FP value for $\tilde\mu$;
with these FP values, one can compute the dimensionless combinations
$\nu$, $\tau$ and $\phi$.
With the spectrally balanced cutoff, there exists a unique value
$\nu_{\rm crit} \approx 0.0956016$,
for which the FP lies exactly on the corresponding hyperbola.
For this choice the chirality condition is UV attractive,
but even in this case the flow does not preserve the hyperbola.

The only point in the $\tilde\Lambda$-$\tilde G$ plane where
$\phi$ becomes approximately RG invariant is near the Gaussian FP,
where the dimensionful $\Lambda$, $G$ and $\mu$ become constant.
In terms of the dimensionless $\tilde\Lambda$ and $\tilde G$
this means that in their beta functions one only keeps the first term on the r.h.s.
of \eq{oneloopbetaasc}.
Then the flow lines approximate the hyperbolas of constant $\tau$
(see right panel in Figure 4).
The approximation becomes better when $\tau$ is small.
If we want to have $\phi=1$ approximately constant this can be achieved
for $\nu=\sqrt{\tau}\to 0$.

\subsection{On-shell properties and the sign of $\Lambda$}

Given that the effective action should be gauge independent on-shell,
i.e. for $R=6\Lambda$, also its $t$-derivative should have this property. From equation \eq{mr}, this implies that $A+6B\tilde\Lambda$ has to be $\alpha$-independent. Indeed one finds that
\be
A+6B\tilde\Lambda=-\frac{16\sqrt{2}}{3\pi}
(1+3\tilde\Lambda)+\textrm{$\mu$-dependent terms}
\ee
so the $\alpha$-dependence cancels.
Another way to see this is to observe that on-shell
the effective action is proportional to $1/\sqrt{\tau}=1/(G\sqrt{\Lambda})$,
up to terms of order $\sqrt{\tilde\Lambda}$.
Thus the beta function for this particular dimensionless combination
of couplings should also be gauge independent.
Indeed, using \eq{oneloopbeta} one finds that this beta function
is proportional to $A+6B\tilde\Lambda$, and therefore $\alpha$-independent.
In view of \eq{relation}, the same is true for the beta function of
$\phi=\mu/\sqrt{|\Lambda|}$.
Note however that the beta functions of $\tau$ and $\phi$
are not functions of $\tau$ and/or $\phi$ alone but depend on
$\tilde\Lambda$ and $\tilde G$ separately. Therefore, to calculate
their flow one needs the beta functions of $\tilde\Lambda$
and $\tilde G$, which have to be computed with off-shell backgrounds.

It is perhaps appropriate to add some further comments on the
significance of the on-shell condition.
One may worry that off-shell backgrounds will give rise to corrections
proportional to the equations of motion. We show in Appendix C
that this is not the case. One may also worry that since the sums \eq{sums} are performed on a three sphere, which has positive curvature, the resulting beta functions are only correct in the domain $\Lambda>0$. In particular, could this not affect the existence of the FP, which actually occurs
(in the descending and balanced schemes) for negative $\Lambda$? Would one not have to perform the calculation on an AdS background?
It would indeed be desirable to perform such a calculation, but since
the Euclidean continuation of AdS spacetime is a noncompact hyperboloid,
the harmonic analysis would be much trickier.
Fortunately there are good reasons to believe that in the UV limit
such a calculation would yield exactly the same beta functions that we have derived here. The reason is that any positive definite metric looks
the same at sufficiently short distance scales, so the beta functions would be the same as long as $k^2\gg R$. Furthermore we observe that for large $\nu$ the actual values of the cosmological constant, in units of $k$, can be made arbitrarily small and negative, while the background can be chosen to have $R$ arbitrarily small and positive, so there exist FPs which are as close to the mass shell as one wishes.


\subsection{Topological issues}


The vanishing of the beta function of the coefficient of the CS term calls
for a topological interpretation.
The analogous phenomenon in gauge theories has been investigated in
\cite{Pisarski:1985yj,Witten:1988hf}.
It was found that in spite of the finiteness
of the theory, the coefficient of the CS term receives a finite renormalization
proportional to the Casimir of the gauge group in the adjoint representation.
This however is an infrared effect \cite{Shifman:1990pa};
in the context of the flow of the effective average action, which we used here,
one can see that
the coefficient of the CS term is constant for all finite $k$ except for
a finite jump in the limit $k\to 0$ \cite{Reuter:1995tr}.
Since our beta functions are only valid in the case $k\gg R$, we do not see
such effect and we cannot exclude that such a finite renormalization happens.

A related point is the following.
It has been argued in \cite{Percacci:1986hu} that as long as one works on a
sphere, it is not necessary to impose quantization of $\nu$, and since we use
precisely a sphere as a background, it is not clear that the vanishing of
the beta function of $\nu$ has a topological explanation.
The answer is that the beta functions we obtain are actually insensitive
to the topology of the manifold and therefore should respect the quantization
of $\nu$ in those cases when the topology demands it.

\subsection{Asymptotic safety and higher derivative terms}

In four dimensional gravity, a nontrivial FP has been found in the leading order
of the $2+\epsilon$ expansion \cite{epsilon} and of a $1/N$ expansion \cite{largen},
in perturbation theory \cite{codello1,niedermaier2} and in a variety of truncations
of an exact RG equation \cite{reuter3,cpr1,bms}
also in the presence of matter \cite{perini}.
The nontrivial FP we have found for each value of $\nu$ in TMG
is similar to the FP that is found in the Einstein-Hilbert
truncation of gravity also in higher dimensions, except for the fact that
the cosmological constant turns out to be negative. There are however two features that make our result particularly interesting.

The first is that for large $\tilde\mu$ the FP value of $\tilde G$ is securely within the
perturbative domain. There is therefore every reason to believe that
the one loop calculation we have performed correctly captures the main features of the Wilsonian flow of this theory. The analogous result in pure three dimensional Einstein Hilbert gravity is physically less appealing because in the absence of CS term the theory does not have propagating degrees of freedom.

The second has to do with the consistency of the truncation.
In principle the effective average action will contain also
higher derivative terms, which in our computation have been neglected.
However, in three dimensions the Riemann tensor is expressed
entirely as a function of the Ricci tensor, and consequently
it is possible to use field redefinitions to eliminate higher derivative terms by means of field redefinitions, order by order in perturbation theory \cite{sen}.
In three dimensions the higher derivative operators are technically redundant and therefore the truncation considered here is consistent. At least in perturbation theory, it is not necessary to consider an infinite set of operators.
However, the redefinitions used to eliminate the higher derivative terms
change the cosmological constant and Newton's constant.
Therefore, the beta functions of $\tilde\Lambda$ and $\tilde G$ will
receive corrections which have not been considered here.


\section*{Acknowledgements}

The work of E.S. is supported in part by NSF grant PHY-0906222. We are grateful to Steven Fulling, Chris Pope, Malcolm Perry and Martin Reuter for helpful discussions.


\newpage

\begin{appendix}

\section{Useful lemmas}

In implementing various cutoff schemes, it is useful to work out separately the decomposition of various terms occurring in the quadratic action
$S^{(2)} + S_{GF}$ given in \eq{a2}. The decomposition of the terms that
come form the Einstein-Hilbert action with a cosmological term are as follows:
\bea
\int d^3 x \sqrt{-g} h_{\mu\nu}\Box h^{\mu\nu} &=& \int d^3 x \sqrt{-g} \Big[ h_{\mu\nu}^T \Box h^{T\mu\nu} -2\xi_\mu \left(\Box + \frac{R}{3}\right) \left(\Box +\frac{2R}{3}\right)\xi^\mu
\nn\w2
&& +\frac{2}{3} \sigma \Box \left(\Box + \frac{R}{2}\right) (\Box + R)\sigma + \frac13 h\Box h\Big]\ ,
\nn\w4
\int d^3 x \sqrt{-g} h_{\mu\nu} \nabla^\mu\nabla_\rho h^{\rho\nu} &=& \int d^3 x \sqrt{-g} \Big[ -\xi_\mu\left(\Box+\frac{R}{3}\right)^2 \xi^\mu +\frac49 \sigma \Box \left(\Box + \frac{R}{2}\right)^2 \sigma
\nn\w2
&& +\frac49 h\Box \left(\Box + \frac{R}{2} \right) \sigma +\frac19 h\Box h\Big]\ ,
\nn\w4
\int d^3 x \sqrt{-g} h \nabla^\mu\nabla^\mu h_{\mu\nu} &=& \int d^3 x \sqrt{-g} \left[ \frac23 h \Box \left(\Box + \frac{R}{2}\right) \sigma +\frac13 h\Box h \right]\ ,
\nn\w4
\int d^3 x \sqrt{-g} h_{\mu\nu} h^{\mu\nu} &=& \int d^3 x \sqrt{-g}
\Big[h_{\mu\nu}^T h^{T\mu\nu} -2\xi_\mu \left(\Box + \frac{R}{3}\right) \xi^\mu
\nn\w2
&& +\frac23 \sigma \Box \left(\Box + \frac{R}{2}\right)\sigma  + \frac13 h^2\Big]\ .
\label{hterms}
\eea

The decomposition of the terms arising in the gravitational CS terms gives

\bea
\int d^3 x h_{\lambda\sigma}\epsilon^{\lambda\mu\nu}\nabla_\mu\Box h_\nu{}^\sigma&=&
\int d^3 x \Big[ h^T_{\lambda\sigma}\epsilon^{\lambda\mu\nu}\nabla_\mu\Box h^T_\nu{}^\sigma
-\xi_\lambda\epsilon^{\lambda\mu\nu}\nabla_\mu \left(\Box+\frac{R}{3}\right)
\left(\Box+\frac{2R}{3}\right)\xi_\nu\Big]
\nn\w4
\int d^3 x h_{\lambda\sigma}\epsilon^{\lambda\mu\nu}\nabla_\mu h_\nu{}^\sigma&=&
\int d^3 x \Big[h^T_{\lambda\sigma}\epsilon^{\lambda\mu\nu}\nabla_\mu h^T_\nu{}^\sigma
-\xi_\lambda\epsilon^{\lambda\mu\nu}\nabla_\mu
\left(\Box+\frac{R}{3}\right)\xi_\nu\Big]
\nn\w4
\int d^3 x h_{\lambda\sigma}\epsilon^{\lambda\mu\nu}\nabla_\mu\nabla^\sigma\nabla^\rho h_{\rho\nu}&=&
\int d^3 x \Big[-\xi_\lambda\epsilon^{\lambda\mu\nu}\nabla_\mu \left(\Box+\frac{R}{3}\right)^2\xi_\nu\Big]\ .
\label{csterms}
\eea

\section{Euclidean continuation and harmonic \hfill\break expansions on $S^3$}

The issue of defining the Euclidean continuation for a theory of gravity is
quite subtle and we will not try to address it here in generality.
In general one would have to think of the real (three dimensional) Minkowskian spacetime as being a section of a three dimensional complex manifold, and find another real section where the metric is positive definite, and where the action is real and possibly bounded from below. The Wick rotation procedure which is employed in perturbative path integral formulation of non-gravitational field theories runs into severe problems in gravity in which the notion of time is affected by diffeomorphisms. Nonetheless, here we will adapt the analog of the standard field theoretic Wick rotation, since we are interested in one-loop beta functions, which are calculable from the gauge fixed quadratic part of the action when expanded about maximally symmetric backgrounds characterized in \eq{sos}. The nature of the Wick rotation still depends on the choice of background metric. For our purposes, it is convenient to work with the $dS_3$ metric, which can be represented as
\be
ds^2=-d\rho^2+\cosh^2\rho(d\theta^2+ \sin^2\theta d\phi^2)\ .
\ee
Upon Wick rotation $\rho\rightarrow -i\rho$, this turns into a metric on $S^3$ given by
\be
ds^2=d\rho^2+\cos^2\rho(d\theta^2+\sin^2\theta d \phi^2)\ .
\ee
Thus in the path integral $e^{iS[g]}$ becomes $e^{-S_E[g_E]}$,
where
\be
S(\bar g)=i \int d\rho d\theta d\phi \sqrt{-\overline g}~{\cal L}[\overline g]
\quad \to \quad
-\int d\rho d\theta d\phi \sqrt{\overline g_E}~{\cal L}[\overline g_E]=-S_E(g_E)\ .
\ee
With this procedure the Euclidean action differs from \eq{1}
just by an overall sign.

To perform harmonic analysis on $S^3$, we decompose all the fields that carry irreducible representations of the stability group $SO(3)\subset SO(4)$, where $SO(4)$ is the isometry group of $S^3$. These fields admit harmonic expansion in terms of all the representation functions of $SO(4)$ that contain the given $SO(3)$ representation (see, for example, \cite{Salam:1981xd}). Thus, the divergence-free and trace-free symmetric tensor $h^T_{\mu\nu}$, the divergence-free vector $\xi$ and a generic scalar $\phi$ are expanded on $S^3$ as
\footnote{We have absorbed the normalization factor $\sqrt{\frac{d(n,s)}{(2j+1)}}$ occurring in the expansion of a spin $j$ field into the definition of the expansion coefficients\cite{Salam:1981xd}, where $d(n,s)$ is the dimension of an $SO(4)$
unitary irreducible representation with highest weight $(n,s)$.}
\bea
h_{\mu\nu}^T (x) &=& \sum_{n=2}^{\infty} \left( h_q^{(n,2)} Y^{(n,2)}_{\mu\nu,q}(x) + h_q^{(n,-2)} Y^{(n,-2)}_{\mu\nu,q}(x)\right)\ ,
\nn\w2
\xi_\mu (x) &=& \sum_{n=1}^{\infty} \left( \xi_q^{(n,1)} Y^{(n,1)}_{\mu,q}(x) + \xi^{(n,-1)} Y^{(n,-1)}_{\mu,q}(x)\right)\ ,
\nn\w2
\phi(x) &=& \sum_{n=0}^{\infty}\phi_q^{(n,0)} Y^{(n,0)}_{0,q}(x) \ ,\eea
where $q$ labels a unitary irreducible representation (UIR) of $SO(4)$, the functions $Y_{\mu\nu,q}$, $Y_{\mu,q}$ and $Y_{0,q}$ are respectively the ${\mu\nu}$'th, $\mu$th and $0$th row and $qth$ column of the Wigner functions for $SO(4)$ UIRs labeled by the highest weight $(n,s)$, with $n\ge |s|$. In expanding a field in spin $j$ representation of $SO(3)$ only the $SO(4)$ UIRs that contain the given $SO(3)$ representation occur, and hence the restriction $n\ge j\ge |s|$. Moreover, the $SO(3)$ representation is contained only once. The expansion coefficients $h^{(n,\pm 2)}$ are constants, and $x$ labels the coordinates on $S^3$. The Wigner functions occurring in these expansions have the same symmetry, trace and divergence properties as the fields that the expanded fields. Note the presence of the zero modes in the form of the single constant mode $Y^{(00)}$, the six Killing vectors $Y_{\mu}^{(1,\pm1)}$ in the adjoint representation of $SO(4)$ and the four conformal Killing vectors $\partial_{\mu}Y^{(10)}$. These obey
\be
\partial_{\mu}Y^{(00)} = 0\ ,\qquad {\overline\nabla}_{\{\mu}{\overline\nabla}_{\nu\}}Y^{(10)} = 0\ ,\qquad
{\overline\nabla}_{(\mu}^{}Y^{(1,\,\pm1)}_{\nu)} = 0\ ,\label{zm}
\ee
where the notation $\{\mu,\nu\}$ means the symmetric and traceless part. The Killing vectors and the conformal Killing vectors together generate
the conformal group ${\rm SO}(4,1)$ that acts on $S^3$.

Next, we make use of the fact that the group theoretical considerations yield the following equations
\bea
-{\overline\Box} Y_{\mu\nu,q}^{(n,\pm 2)}(x) &=& {R\over 6} [(n+1)^2-3]\,Y_{\mu\nu,q}^{(n,\pm 2)}(x)\ ,
\w2
-{\overline\Box} Y_{\mu,q}^{(n,\pm 1)}(x) &=& {R\over 6} [(n+1)^2-2]\,Y_{\mu,q}^{(n,\pm 1)}(x)\ ,
\w2
{\overline\nabla}_{[\mu} Y_{\nu],q}^{(n,\pm 1)}(x) &=& \pm\, \frac12 \sqrt{\frac{R}{6}} (n+1)
 \bar \varepsilon_{\mu\nu}{}^\rho \,Y_{\rho,q}^{(n,\pm 1)}(x)\ ,
\w2
-{\overline\Box} Y_{0,q}^{(n,0)}(x) &=& {R\over 6} [(n+1)^2-1]\,Y_{0,q}^{(n,0)}(x)\ ,
\w2
{\overline\nabla}_{[\mu} Y_{\nu]\rho,q}^{(n,\pm 2)}(x) &=& \pm\, i\sqrt{\frac{R}{6}} (n+1)
 \bar \varepsilon_{\mu\nu}{}^\sigma \,Y_{\rho\sigma,q}^{(n,\pm 2)}(x)\ .
\eea
with multiplicities as given in \eq{multiplicities}. In deriving these formulae, we use the well known formulae for the Casimir eigenvalues of the second-order Casimir operator $C_2(n,s)$ for an $SO(4)$ representation labeled by $(n,s)$, and it dimension $d(n,s)$ given by
\be
C_2(n,s) =  n(n+2) +s^2\ ,   \qquad d(n,s)= (n-1)^2 -s^2\ .
\ee
Furthermore, we use the formula
\be
-{\overline\Box} Y_{j,q}^{(n,s)}(x) = \frac{R}{6} \left[ C_2(n,s)-C_2(j)\right] Y_{j,q}^{(n,s)}(x)\ ,
\ee
where $Y_{(j) q}^{(n,s)}(x)$ is the Wigner function for the $SO(4)$ UIR in the $(n,s)$ representation, viewed as a matrix, restricted in its row to the lowest spin $j$ representation it contains, and $C_2(j)=j(j+1)$ is the eigenvalue of the second-order Casimir operator for $SO(3)$ in spin $j$ representation.

\section{The off shell one loop beta functional}

In the evaluation of the one loop effective action using the background field method one
usually assumes that the background is a solution of the classical equations of motion.
On the other hand in the derivation of the beta functions for gravity we have to evaluate
$\partial_t \Gamma_k$ for off-shell backgrounds. This is necessary in order to
disentangle the beta functions of $G$ and $\Lambda$.
One may worry that for off-shell backgrounds the simple equation \eq{oneloop}
is no longer correct. We show here that this is not the case.

In order to do this, we review the derivation of that equation within the
background field method.
We consider a classical action $S(\phi)$ and expand $\phi=\bar\phi+\eta$
where the background $\bar\phi$ is arbitrary. We Taylor expand
\be
S(\phi)=S(\bar\phi)+\int \frac{\delta S}{\delta\phi}\Big|_{\bar\phi} \eta
+\frac{1}{2}\int\!\!\int\eta \Delta(\bar\phi) \eta+\ldots
\ee
where
$\Delta_{\bar\phi}=\frac{\delta^2 S}{\delta\phi\delta\phi}\Big|_{\bar\phi}$.
The $k$-dependent generating functional analogous to \eq{ge} can now be written as

\begin{equation}
e^{-W_{k}\left[\bar\phi,j\right]}=
\int D\eta\exp\left\{-S[\bar\phi+\eta]-\Delta S_{k}[\eta]-\int \, j\eta\right\}\ ,
\end{equation}
such that
\be
\label{onepoint}
\varphi\equiv\langle\eta\rangle=\frac{\delta W_k}{\delta j}\ .
\ee
The background-dependent effective average action is the $k$-dependent modified Legendre transform
\begin{equation}
\label{legendre}
\Gamma_{k}[\bar\phi,\varphi]=W_{k}\left[j\right]-\int \, j\varphi-\Delta
S_{k}[\varphi]\,,
\end{equation}
where $j$ has to be expressed as a function of $\varphi$, solving \eq{onepoint}.
From here one finds that
\begin{equation}
\label{cleq}
\frac{\delta\Gamma_{k}}{\delta\varphi}=-j-R_k\varphi\ ;\qquad
\frac{\delta^2\Gamma_{k}}{\delta\varphi\delta\varphi}=
-\left(\frac{\delta^2 W_k}{\delta j\delta j}\right)^{-1}-R_k\
\,.
\end{equation}

Let us evaluate these functionals at one loop.
We have
\be
W_k^{(1)}[\bar\phi,j]=S(\bar\phi)
+\frac{1}{2}\mathrm{Tr\,log}\left(\Delta_{\bar\phi}+R_k(\Delta_{\bar\phi})\right)
-\frac{1}{2}\int\!\!\int j(\Delta_{\bar\phi}+R_k(\Delta_{\bar\phi}))^{-1}j\ .
\ee
From here solving equation \eq{onepoint} we get
\be
j=-(\Delta_{\bar\phi}+R_k(\Delta_{\bar\phi}))\varphi\,.
\ee
Therefore the one loop average effective action $\Gamma_k^{(1)}$ is given by
\be
\label{ansatz}
\Gamma^{(1)}_k(\bar\phi,\varphi)
=S(\bar\phi)+\frac{1}{2}{\rm Tr}\,{\rm log}\left(\Delta_{\bar\phi}+R_k(\Delta_{\bar\phi})\right)
+\frac{1}{2}\int\!\!\int \varphi \Delta_{\bar\phi} \varphi\ .
\ee
The main point to observe here is that the terms of the form $\int\varphi R_k\,\varphi$
have exactly cancelled.
Therefore, the only dependence on $k$ is in the trace term, and the beta functional
\eq{oneloop} is unaffected. In a gauge theory, the presence of the gauge fixing terms does not
change this conclusion.

\section{Gravity without CS term}

Here we derive the beta functions of $\tilde\Lambda$ and $\tilde G$
using heat kernel techniques.
This has been described elsewhere in a variety of gauges and cutoff schemes,
but not using the cutoff scheme described in Section 3.
For the sake of comparison with the beta functions given in Section 5
we give here this calculation.

In the absence of CS term, and in diagonal gauge, all the operators appearing in the functional traces \eq{erge3} are minimal Laplace-type operators of the form $\Delta=-\nabla^2\mathbf{1}+\mathbf{E}$
where $\mathbf{E}$ is a linear map acting on the spacetime and internal indices of the fields.
In our applications it will have the form
$\mathbf{E}=(q_1 R+q_2\Lambda)\, {\bf 1}+$ where $\mathbf{1}$ is the identity in the space of the fields
and $q_1$, $q_2$ are real numbers.

The trace of a function $W$ of the operator $\Delta$ can be written as
\begin{equation}
\label{c1}
{\rm Tr}W(\Delta)=\sum_i W(\lambda_i)=\int_0^\infty ds {\rm Tr}K(s)\tilde W(s)
\end{equation}
where $\lambda_i$ are the eigenvalues of $\Delta$,
$\tilde W$ is the Laplace anti-transform of $W(s)$
and ${\rm Tr}K(s)=\sum_i e^{-s\lambda_i}$ is the trace of the heat kernel of $\Delta$.
The UV behavior of the theory is governed by the lower end of the integration over $s$.
We use the asymptotic expansion for $s\to 0$:
\be
\label{heatkernel}
{\rm Tr} \left(e^{-s\Delta}\right)=
\frac{1}{(4\pi)^{d/2}}
\left[B_0(\Delta)s^{-\frac{d}{2}}+B_2(\Delta)s^{-\frac{d}{2}+1}
+\ldots+B_d(\Delta)+B_{d+2}(\Delta)s+...\right]
\ee
where $B_n=\int{\rm d}^d\, x\sqrt{g}{\rm tr}\mathbf{b}_n$ and $\mathbf{b}_n$ are linear combinations
of curvature tensors and their covariant derivatives containing $n$ derivatives of the metric.
Then \eq{c1} becomes
\begin{align}
\label{HKasymp}
{\rm Tr}W(\Delta)=\frac{1}{(4\pi)^{d/2}}
\bigl[&Q_   {\frac{d}{2}}(W)B_0(\Delta)+Q_{\frac{d}{2}-1}(W)B_2(\Delta)+\ldots\nonumber\\
&+Q_0(W)B_{d}(\Delta)+Q_{-1}(W)B_{d+2}(\Delta)
+\ldots\bigr]\ ,
\end{align}
where
\begin{equation}
\label{q1}
Q_n(W)=\int_0^\infty ds s^{-n}\tilde W(s)=\frac{1}{\Gamma(n)}\int_{0}^{\infty}dz\, z^{n-1}W(z)
\ .
\end{equation}
The last equality only holds for $n>0$.

For the function $W=\frac{\partial_t R_k}{P_k}$ of interest in the evaluation of \eq{erge3} the integrals are
\begin{equation}
\label{q1}
Q_{3/2}(W)=\frac{8}{3\sqrt{\pi}}k^3  \ ;\qquad
Q_{1/2}(W)=\frac{4}{\sqrt{\pi}}k
\end{equation}

The only remaining ingredient that is needed for the evaluation of \eq{mr} are the coefficients $B_0$ and $B_2$ of the operators $\Delta$
given in \eq{ops}.
These can be evaluated using standard methods (see for example Appendix B in \cite{cpr2}) and are listed in the table.

\begin{table}
\begin{center}
\begin{tabular}{|c|c|c|c|c|c|c|}\hline
  & $h^T$ & $\xi$ & $\sigma$ & $h$ & $V$ & $S$ \\\hline
$\mathrm{tr}\mathbf{b}_0$ & $2$ & $2$ & $1$ & $1$ & $2$ & $1$ \\\hline
$\mathrm{tr}\mathbf{b}_2$ &\! $-3R+4\Lambda$ & $\frac{2(1-\alpha)R}{3}+4\alpha\Lambda$ &
$\frac{(16-7\alpha)R}{6(4-\alpha)}+\frac{6\alpha\Lambda}{4-\alpha}$ &
$\frac{(10-\alpha)R}{6(4-\alpha)}+\frac{6\Lambda}{4-\alpha}$  &
$\frac{2}{3}$ &
$\frac{16-\alpha}{6(4-\alpha)}R$   \\\hline
\end{tabular}\end{center}
\end{table}

The result is
\be
\partial_t\Gamma_k=\int d^3x\sqrt{g}\,\frac{6(11-9\alpha-2\alpha^2)\Lambda-(47-2\alpha^2) R}{12\pi^2(4-\alpha)}k\ ,
\ee
from which we read off
\be
A= \frac{8(11+9\alpha-2\alpha^2)\tilde\Lambda}{\pi(4-\alpha)}  \ ;\qquad
B= -\frac{4(47-2\alpha^2)}{3\pi(4-\alpha)}  \ .
\ee
These results agree, for $\alpha=1$,
with those reported in Section IV D of \cite{cpr2}.
As discussed there, the values of $A$ and $B$ do depend on the cutoff scheme.
This is normal for the beta functions of dimensionful couplings.
It is reassuring that this scheme dependence does not change the picture qualitatively. Furthermore, there are certain aspects of the flow that are scheme independent. The first one is the beta function of the dimensionless coupling $\nu$, which is zero independently on the scheme. Another one is the beta function of the dimensionless combination $\tau$: it is equal to
\be
\partial_t (\Lambda G^2)=-\frac{36\tilde\Lambda\tilde G^3}{\pi}
\ee
in both cases. This accords with the gauge independence of the same quantity, as discussed in Section 6. Higher terms of the expansion in $\tilde\Lambda$ will probably not be
likewise universal.

We conclude by giving the properties of the nontrivial FP in this scheme.
It occurs at $\tilde\Lambda=0$, $\tilde G=0.2005$; its scaling exponents
are $-1$ and $-2.298$, with eigenvectors aligned with the
coordinate axes.

\end{appendix}

\end{document}